\documentclass[preprint,pre,longbibliography,eqsecnum]{revtex4-1} 

\usepackage{natbib}
\usepackage{graphicx}
\usepackage{subfigure}
\usepackage{amssymb}
\usepackage{amsfonts}
\usepackage{amsmath}
\usepackage{amsbsy}
\usepackage{mathrsfs} 
\usepackage{color}
\usepackage{hyperref} 
\usepackage{soul}
\usepackage{mymacros}

\begin{document} 

\begin{abstract}
We develop a reduced model for the slow unsteady dynamics of an isotropic chemically active particle near the threshold for spontaneous motion. Building on the steady theory developed in part I of this series, we match a weakly nonlinear expansion valid on the particle scale with a leading-order approximation in a larger-scale unsteady remote region, where the particle acts as a moving point source of diffusing concentration. The resulting amplitude equation for the velocity of the particle includes a term representing the interaction of the particle with its own concentration wake in the remote region, which can be expressed as a time integral over the history of the particle motion, allowing efficient simulation and theoretical analysis of fully three-dimensional unsteady problems. To illustrate how to use the model, we study the effects of a weak force acting on the particle, including the stability of the steady states and how the velocity vector realigns towards the stable one, neither of which previous axisymmetric and steady models were able to capture. This unsteady formulation could also be applied to most of the other perturbation scenarios studied in part I as well as the dynamics of interacting active particles. 
\end{abstract}

\title{\normalsize Weakly nonlinear dynamics of a chemically  active particle near the threshold for  spontaneous motion. 
II. History-dependent motion}

\author{Gunnar G. Peng}
\author{Ory Schnitzer}
\affiliation{Department of Mathematics, Imperial College London, London SW7 2AZ, UK}

\maketitle
\section{Introduction}\label{sec:introduction}
This is the second part of a series of papers developing a weakly nonlinear theory for the dynamics of chemically active particles near the threshold for spontaneous motion. The starting point for the theory is the canonical model proposed by Michelin \textit{et al.} \cite{Michelin:13} describing an autocatalytic colloid that self-propels despite having uniform surface properties, the underlying mechanism being a hydro-chemical instability that can occur when diffusion is sufficiently weak relative to advection. While in practice diffusion dominates in experiments involving autocatalytic colloids \cite{Golestanian:05,Golestanian:07}, the canonical active-particle model serves as a simplified yet physically consistent reference model for self-solubilizing chemically active drops  whose spontaneous motion has been observed in many experiments (see the recent review \cite{Michelin:22} and references therein). As discussed in the first part of this series \cite{Schnitzer:22} (henceforth `part I'), weakly nonlinear analyses of the  canonical active-particle model and similar models of self-solubizing  active drops have previously been limited to steady, axisymmetric solutions \cite{Rednikov:94,Rednikov:94b,Morozov:19,Morozov:19b,Saha:21}. Building on those works, this series aims at an unsteady, three-dimensional theory that is also versatile in the sense that various perturbation effects can be easily included.  

In part I, we revisited the weakly nonlinear analysis of the canonical model in the steady case. The key contribution of that part was the identification of the adjoint linearized differential operator and auxiliary conditions at the threshold. This allowed us to circumvent the apparent need in previous studies to directly solve the inhomogeneous linear problem at second order of the weakly nonlinear expansion. Rather, we used a Fredholm Alternative argument to extract a solvability condition on that inhomogeneous problem directly from its formulation, thereby obtaining the nonlinear amplitude equation governing the steady-state particle velocity. While this contribution may seem merely technical, it in fact enables significant  simplification and generalization. In particular, it effectively eliminates the difficulty in analyzing three-dimensional (in contrast to collinear/axisymmetric) particle motion and makes it straightforward to study the influence of weak perturbations from the canonical model, which generally have a leading-order effect on the particle velocity sufficiently near the threshold. We extensively demonstrated the efficacy of this approach by showing how the singular-pitchfork bifurcation previously established for isotropic active particles \cite{Morozov:19,Morozov:19b,Farutin:21,Saha:21} is modified when various perturbations are included such as external force or torque fields, non-uniform surface perturbations and surface and bulk chemical kinetics. 

In this part, we shall extend the steady  weakly nonlinear theory developed in part I to the unsteady case. The motivation for such an extension is clear, namely to allow one to study the stability of steady states, transients and explore more complex unsteady phenomena that may arise when various perturbation effects are included into the modelling. 

Following the  steady case, our analysis will be based on asymptotic matching of a particle-scale weakly nonlinear expansion with a leading-order approximation in a remote region describing the concentration wake associated with the particle's activity and motion. We will see that the particle velocity evolves on such a long time scale that the particle-scale weakly nonlinear expansion remains quasi-steady at the relevant orders. As as consequence, we will be able to directly employ the particle-scale analysis from part I,  including the adjoint formulation associated with that region. In contrast, the remote region will be seen to be unsteady already at leading order, requiring us to generalize the analysis of that region and its matching with the particle-scale region.

As anticipated in part I, the above picture will lead us to a non-conventional manifestation of unsteadiness in the nonlinear amplitude equation for the particle velocity, namely as an integral over the history of the particle motion representing the interaction of the particle with its own wake. This history dependence is linked to the spatial non-uniformity of the weakly nonlinear expansion, much like the other, more familiar, unconventional features of this problem, such as the singular-pitchfork bifurcation in the unperturbed isotropic active-particle case and the fact that the amplitude equation arises from solvability at second, rather than third, order of the weakly nonlinear expansion \cite{Morozov:19,Morozov:19b,Saha:21,Farutin:21}. 

To focus on this essentially novel aspect of the theory, the only perturbation to the canonical active-particle model we shall consider in this part is that of a weak (possibly time-dependent) external force. There are two good reasons for considering this particular perturbation. First, the problem of an isotropic active particle or drop in an external force field has received considerable analytical \cite{Rednikov:94,Rednikov:94b,Yariv:17:Uri,Saha:21}, numerical \cite{Kailasham:22} and experimental attention \cite{Castonguay:22}. We note that weakly nonlinear analysis has been employed previously in order to calculate steady states in this scenario, in \cite{Saha:21} and part I for the canonical active-particle model and before that in \cite{Rednikov:94,Rednikov:94b} for closely related active-drop models. Their stability, however, has so far only been guessed \cite{Saha:21} or heuristically argued \cite{Rednikov:94,Rednikov:94b}. Second, it will be convenient to demonstrate stability predictions by considering the dynamics of particles disturbed from steady-state motion (force-free or forced) by localized-in-time force perturbations. 

This paper continues as follows. In Sec.~\ref{sec:formulation}, we formulate the canonical active-particle model with unsteadiness, and including a time-dependent external force. In Sec.~\ref{sec:wna}, we develop the unsteady weakly nonlinear theory building on the steady analysis of part I. The analysis in this section furnishes a preliminary form of the amplitude equation for the particle velocity wherein unsteadiness enters via coupling with an initial-boundary-value problem in the remote region. In Sec.~\ref{sec:history}, we show by analysis of that problem that unsteadiness can instead be represented by a nonlocal history operator, namely a time integral over the history of the particle motion. In Sec.~\ref{sec:dynamics}, we present theoretical and numerical predictions of our model for both force-free and forced particles. We give concluding remarks in Sec.~\ref{sec:conclusions}. 

\section{Problem formulation}
\label{sec:formulation}
\subsection{Isotropic chemically active particle with an external force}
We adopt the canonical model of an isotropic chemically active particle proposed by Michelin \textit{et al.} \cite{Michelin:13}, except that following several subsequent studies we include the possibility of an external force  \cite{Yariv:17:Uri,Saha:21,Kailasham:22}. The formulation is thus similar to that in part I of this series, specifically in the scenario where an external force is included, with the key difference that in this part we allow for unsteadiness. 

Consider a homogeneous spherical particle of radius $a_*$ immersed in an unbounded liquid of viscosity $\eta_*$, with an asterisk subscript indicating a dimensional quantity. Solute molecules dissolved in the liquid are transported by diffusion, with diffusivity $D_*$, as well as advection, with the concentration approaching a constant value far from the particle. (Only perturbations from this value will be important.) 
The chemical activity of the particle is represented by a uniform and constant solute flux $j_*$ at the particle surface (positive into the liquid). The fluid is driven at the surface of the particle via a diffusio-osmotic slip mechanism such that the fluid velocity relative to the surface is locally proportional to the concentration surface gradient; the coefficient of proportionality $b_*$, namely the slip coefficient, or surface mobility, is uniform and constant. 

It is assumed that the inertia of the particle and fluid can be neglected. The flow field is therefore governed by the Stokes (zero-Reynolds-number) limit of the Navier--Stokes equations and the motion of the particle is such that the net (i.e., hydrodynamic plus external) force and torque on the particle both vanish. In this study, we shall only include an external force, denoted by $\mathbf{F}\ub{e}_*(t_*)$, which depends on time $t_*$. 

In the case where there is no force, it is known that instability and spontaneous motion occur only if $j_*b_*>0$ \cite{Michelin:13}. This is also a necessary condition for enhanced translation in the case where there is a weak external force  \cite{Saha:21,Kailasham:22}. Without loss of generality, we shall assume that both $j_*$ and $b_*$ are positive.

\subsection{Dimensionless formulation}
\begin{figure}[t!]
\begin{center}
\includegraphics[scale=0.5]{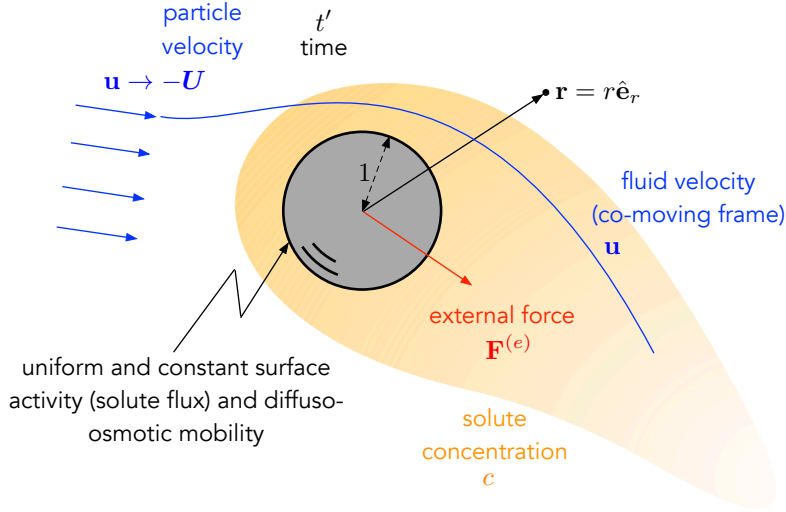}
\caption{Dimensionless schematic of a forced isotropic active particle. The angular velocity of the particle turns out to vanish trivially.}
\label{fig:sketch}
\end{center}
\end{figure}

We adopt the same dimensionless conventions as in part I: lengths are normalized by $a_*$, concentrations by $c_* = a_*j_*/D_*$, velocities by $u_*=j_*b_*/D_*$, stresses and pressures by $\eta_*u_*/a_*$, forces by $\eta_*u_*a_*$ and torques by $\eta_*u_*a_*^2$. Time is normalized by $a_*/u_*$. As shown in Fig.~\ref{fig:sketch}, we denote the corresponding position vector measured from the center of the particle by $\br=r\be_r$, with $r=|\br|$ and $\be_r$ a radial unit vector; the deviation of the solute concentration from its value at infinity by $c$ (henceforth called concentration); the fluid velocity field in a non-rotating frame of reference that moves with the particle's centroid by $\bu$; the associated pressure field by $p$; the centroid velocity by $\bU$; and particle angular velocity by $\boldsymbol{\Omega}$. We denote time by $t'$, reserving the usual notation $t$ for a slow-time coordinate to be introduced in the next section for the purpose of a weakly nonlinear analysis. The external force is denoted by $\mathbf{F}\ub{e}$. 

The concentration $c$ satisfies the unsteady advection--diffusion equation
\begin{equation}\label{c eq}
\Pen\,\left(\pd{c}{t'}+\bu\bcdot\bnabla c\right)=\nabla^2c,
\end{equation}
where
\begin{equation}\label{Pe def}
\Pen=\frac{a_*u_*}{D_*}
\end{equation}
is an `intrinsic' P\'eclet number measuring the importance of advection relative to diffusion; 
the boundary condition 
\begin{equation}\label{c bc}
\pd{c}{r}=-1 \quad \text{at} \quad r=1
\end{equation}
and decay condition
\begin{equation}\label{c far}
c\to0 \quad \text{as} \quad r\to\infty.
\end{equation}
The flow $\bu$ and pressure $p$ satisfy the Stokes equations 
\refstepcounter{equation}
$$
\label{u eqs}
\bnabla \bcdot \bu  = 0, \quad \bnabla\bcdot\boldsymbol{\sigma}=\bzero,
\eqno{(\theequation a,b)}
$$
in which 
\begin{equation}\label{stress}
\boldsymbol{\sigma}=-p\tI + \bnabla\bu + (\bnabla\bu)^\dagger
\end{equation} 
is the stress tensor, $\tI$ being the identity tensor and $\dagger$ denoting the tensor transpose; 
the boundary condition
\begin{equation}\label{u bc}
\bu=\bnabla_s c +\mathbf{\Omega}\times \br \quad \text{at} \quad r=1,
\end{equation}
in which $\bnabla_s=(\tI-\be_r\be_r)\bcdot\bnabla$ is the surface-gradient operator; 
the far-field condition
\begin{equation}\label{u far}
\bu \to -\mathbf{U} \quad \text{as} \quad r\to\infty;
\end{equation}
and the integral conditions
\refstepcounter{equation}
$$
\label{force torque balances}
\bF+\bF\ub{e}= \bzero, \quad \bT=\bzero,
\eqno{(\theequation a,b)}
$$
where the hydrodynamic force $\bF$ and torque $\bT$ are defined by 
\refstepcounter{equation}
$$
\label{force torque}
\bF= \oint_{r=1} \mathrm{d}A\, \be_r\bcdot \boldsymbol{\sigma}, \quad \bT= \oint_{r=1} \mathrm{d}A\, \br\times (\be_r\bcdot \boldsymbol{\sigma}),
\eqno{(\theequation a,b)}
$$
wherein $\mathrm{d}A$ is an infinitesimal area element. As the Stokes equations \eqref{u eqs} are homogeneous, the pressure $p$ should be interpreted as a modified pressure with any deviation from neutral buoyancy of the particle included in the external force. 

In principle, it is necessary to also prescribe initial conditions in order to close the problem formulation. Since the flow problem is instantaneous, it suffices to prescribe the concentration field at any given moment. We shall return to this point in the next section. 

For the problem formulated above, it can be shown that the angular velocity $\boldsymbol{\Omega}$ vanishes trivially. This follows from an application of the reciprocal theorem to spherical phoretic particles as in \cite{Stone:96}, noting the absence of an external torque (cf.~(\ref{force torque balances}b)) and the fact that the relative surface velocity is given by the surface gradient of a smooth scalar field (cf.~\eqref{u bc}). We shall not rely on this exact result so that the analysis in the next section appears as close as possible to that in part I. This will help to simplify future extensions of the unsteady analysis in this part to include  perturbation effects that give rise to particle rotation.

\section{Unsteady weakly nonlinear theory}\label{sec:wna}
\subsection{Weakly nonlinear regime and slow time scale}
In the absence of an external force, there exists a steady solution for all values of the P\'eclet number such that the particle and fluid are stationary and the concentration is spherically distributed, $c=c_0(r)$, with
\begin{equation}\label{c0}
c_0=\frac{1}{r}.
\end{equation} 
The linear stability analysis in Ref.~\cite{Michelin:13}, which assumes axisymmetric perturbations, shows that this stationary--symmetric base state is unstable for $\Pen>4$, with a positive growth rate (normalized by $u_*/a_*$) scaling like $(\Pen-4)^2$ as $\Pen\searrow4$. This scaling suggests that near the instability threshold the dynamics of the particle evolve on long  times scaling like $1/(\Pen-4)^2$, the physical significance of which will become evident later. It is further known, based on the steady analyses in part I and Ref.~\cite{Saha:21}, that even a weak external force can have a leading-order effect on the motion of the particle sufficiently near the threshold, specifically for $\Pen-4 = \mathcal{O}(|\mathbf{F}\ub{e}|^{1/2})$.

Our goal here is to describe the unsteady dynamics of the particle near the instability threshold, including the effect of a weak external force. As in part I, we write 
\begin{equation}\label{chi def}
\Pen=4 + \epsilon \chi
\end{equation}
and carry out a weakly nonlinear analysis as $\epsilon\searrow0$, with  $\chi$ a rescaled bifurcation parameter. In light of the above scalings, the concentration and flow fields are assumed to be functions of position $\br$ and the slow time coordinate
\begin{equation}\label{slow time}
t=\epsilon^2t',
\end{equation}
such that, in particular, the advection--diffusion equation \eqref{c eq} is rewritten as
\begin{equation}\label{AD T}
\Pen\,\left(\epsilon^2\pd{c}{t}+\bu\bcdot\bnabla c\right)=\nabla^2c.
\end{equation}
Furthermore, the external force is prescribed as
\begin{equation}
\mathbf{F}\ub{e}(t')=6\pi \epsilon^2 \mathbf{f}(t),
\end{equation}
so that it varies on the characteristic time scale and is just strong enough to have a leading-order effect on the particle motion.

\begin{figure}[t!]
\begin{center}
\includegraphics[scale=0.4]{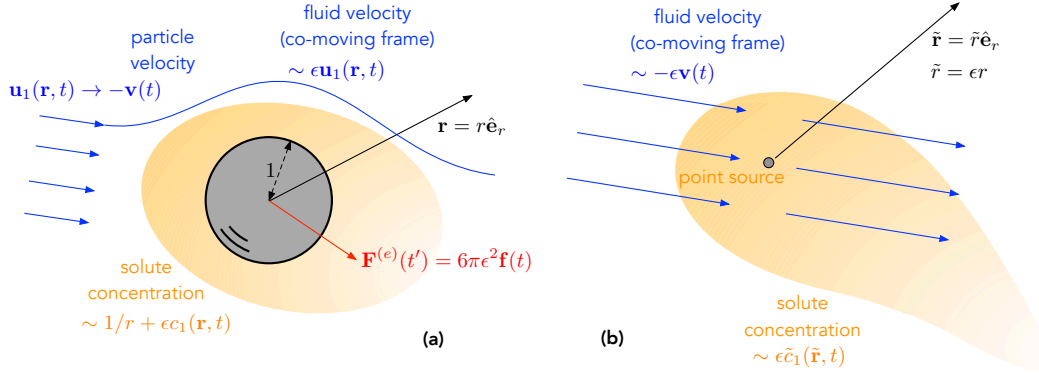}
\caption{Particle-scale (a) and remote (b) regions of the weakly nonlinear analysis near the instability threshold  (Sec.~\ref{sec:wna}). The small positive parameter $\epsilon$ characterizes the shift of the P\'eclet number from its critical value (cf.~\eqref{chi def}). Note that $t=\epsilon^2t'$ denotes a slow-time coordinate (cf.~\eqref{slow time}).} 
\label{fig:sketchWNA}
\end{center}
\end{figure}

\subsection{Particle-scale weakly nonlinear expansions}
We expand the concentration in powers of $\epsilon$ as
\begin{equation}\label{c expansion}
c(\br,t;\epsilon) \sim c_0(r)+\epsilon c_1(\br,t)+\epsilon^2 c_2(\br,t) \quad \text{as} \quad \epsilon\searrow 0,
\end{equation}
wherein $c_0$ is the equilibrium solution \eqref{c0}, and the flow as
\refstepcounter{equation}
$$
\label{u p expansions}
\bu(\br,t;\epsilon) \sim \epsilon\bu_{1}(\br,t) + \epsilon^2 \bu_2(\br,t), \quad p(\br,t;\epsilon)\sim \epsilon p_{1}(\br,t)+\epsilon^2 p_2(\br,t) \quad \text{as} \quad \epsilon\searrow0.
\eqno{(\theequation a,b)}
$$
Quantities associated with the flow, including the hydrodynamic force $\mathbf{F}$, hydrodynamic torque $\mathbf{T}$, stress tensor $\boldsymbol{\sigma}$, and particle velocities $\bU$ and $\boldsymbol{\Omega}$, are expanded similarly to \eqref{u p expansions}. For the particle velocity $\bU$, we introduce the special notation $\bU_1(t)=\bv(t)$, such that
\begin{equation}\label{v def}
\bU(t;\epsilon) \sim \epsilon \bv(t) \quad \text{as} \quad \epsilon\searrow0,
\end{equation}
with $\bv(t)$ corresponding to a rescaled leading-order approximation for the particle velocity. Given prior comments, we expect the expansion for $\boldsymbol{\Omega}$ to be trivial at all orders. 

For the same reasons as discussed in part I, the concentration `particle-scale' expansion \eqref{c expansion}  breaks down at large distances $r=\text{ord}(1/\epsilon)$  corresponding to a `remote region' where diffusion and advection are comparable and in particular $c_0=1/r$ does not hold as a leading-order approximation (see Fig.~\ref{fig:sketchWNA}). 

In the particle region, $c_0$ is independent of $t$ and so the time derivative in \eqref{AD T} does not enter up to and including the order-$\epsilon^2$ balance of that equation. Based on the steady analysis in part I, we do not expect higher-order terms to influence the leading-order dynamics of the particle. Thus, the particle-scale region is effectively quasi-static and hence its  analysis follows closely the steady analysis in part I. In particular, we will be able to directly employ the solvability result derived in part I in order to circumvent the need to solve the particle-scale problem at order $\epsilon^2$. The only difference arises from the far-field conditions satisfied by $c_1(\br,t)$ and $c_2(\br,t)$ as $r\to\infty$, which now need to be derived by matching with a generally unsteady, rather than steady, remote-region approximation. 

We proceed with an analysis of the unsteady remote region (Sec.~\ref{ssec:remote}), after which it will be straightforward to utilize results from part I pertaining to the particle-scale expansions \eqref{c expansion} and \eqref{u p expansions} in order to derive an unsteady nonlinear amplitude equation for $\bv(t)$  (Sec.~\ref{ssec:amplitude}).

\subsection{Unsteady remote region}\label{ssec:remote}
Towards analyzing the remote region, we define $c(\br,t;\epsilon)=\tilde{c}(\tilde{\br},t;\epsilon)$, where $\tilde{\br}=\epsilon \br$ is a strained position vector of magnitude $\tilde{r}=\epsilon r$. As in the steady analysis of part  I, the $1/r$ decay of the leading-order particle-scale concentration (cf.~\eqref{c0}) implies the remote-region expansion 
\begin{equation}\label{remote expansion}
\tilde{c}(\tilde{\br},t;\epsilon) \sim \epsilon \tilde{c}_1(\tilde{\br},t) \quad \text{as} \quad \epsilon\searrow 0.
\end{equation}
In contrast to the particle-scale concentration expansion \eqref{c expansion}, the flow expansions \eqref{u p expansions} are uniform in $r$. Accordingly, the far-field condition \eqref{u far} together with the leading-order approximation \eqref{v def} for the particle velocity together imply $\bu(\tilde{\br}/\epsilon,t;\epsilon)\sim -\epsilon \bv(t)$ as $\epsilon\searrow0$. Thus, together with the P\'eclet-number rescaling \eqref{chi def}, the advection--diffusion equation \eqref{AD T} gives 
\begin{equation}\label{remote eq} 
\pd{\tilde{c}_1}{t}-\bv \bcdot \tilde{\bnabla}\tilde{c}_1=\frac{1}{4}\tilde{\nabla}^2 \tilde{c}_1,
\end{equation}
for $\tilde{r}>0$, 
in which $\tilde{\bnabla}$ is the gradient operator with respect to $\tilde{\br}$. The advection--diffusion equation \eqref{remote eq}, in which the advecting flow is a time-dependent uniform stream, is supplemented by the decay condition
\begin{equation}\label{remote far}
\tilde{c}_1\to0\quad\text{as} \quad \tilde{r}\to\infty,
\end{equation}
which follows from \eqref{c far}, and the singular behavior
\begin{equation}\label{remote matching}
\tilde{c}_1\sim \frac{1}{\tilde{r}}\quad \text{as} \quad \tilde{r}\searrow0,
\end{equation}
which follows from asymptotically matching the one-term remote-region expansion \eqref{remote expansion} and the particle-region concentration expansion \eqref{c expansion} taken to leading order. 

The problem \eqref{remote eq}--\eqref{remote matching} governing $\tilde{c}_1(\br,t)$ is schematically depicted in Fig.~\ref{fig:sketchWNA}b. It is an unsteady generalization of the steady remote-region problem encountered in part I. For the purpose of deriving far-field conditions on the particle-scale concentrations $c_1(\br,t)$ and $c_2(\br,t)$, what is important is the behavior of $\tilde{c}(\tilde{\br},t)$ as $\tilde{\br}\searrow0$. In appendix \ref{app:localanalysis}, we show by means of a local analysis of \eqref{remote eq} as $\tilde{r}\searrow0$, starting with \eqref{remote matching}, that the solution possesses an expansion of the form
\begin{multline}\label{local behaviour}
\tilde{c}_1(\tilde{\br},t)=\frac{1}{\tilde{r}}+\tilde{h}(t)-2\bv(t)\bcdot\be_r\\+\tilde{r}\left\{2\bv(t)\bv(t)\boldsymbol{:}(\tI+\be_r\be_r)+\be_r\bcdot\tilde{\mathbf{H}}(t)\right\} + o(\tilde{r}) \quad \text{as} \quad \tilde{r}\searrow0,
\end{multline}
in which $\tilde{h}(t)$ and $\tilde{\mathbf{H}}(t)$ are a scalar and a vector, respectively, that are left undetermined by the local analysis. Given a global, i.e., exact solution to the remote-region problem \eqref{remote eq}--\eqref{remote matching}, these functions could be extracted at any  time $t$ by inspecting the behavior of the solution as $\tilde{r}\searrow0$. We shall see that $\tilde{\mathbf{H}}(t)$ plays a role in determining the leading-order dynamics of the particle, whereas $\tilde{h}(t)$ does not.

In particular, for a steady particle velocity $\bv(t)=\bar{\bv}$ having magnitude $\bar{v}$, the remote-region problem \eqref{remote eq}--\eqref{remote matching} is known to possess the steady solution \cite{Acrivos:62}
\begin{equation}\label{remote sol steady}
\tilde{c}_1\ub{s}(\tilde{\br};\bar{\bv})=\frac{1}{\tilde{r}}\exp\left\{-2\bar{\bv}\bcdot \tilde{\br}-2\bar{v}\tilde{r}\right\},
\end{equation}
which was also used in steady weakly nonlinear analyses of active particles \cite{Morozov:19,Morozov:19b,Saha:21}. It has the local expansion
\begin{equation}\label{remote sol steady expansion}
\tilde{c}_1\ub{s}(\tilde{\br};\bar{\bv})= \frac{1}{\tilde{r}}-2\bar{v}-2\bar{\bv}\bcdot\be_r +\tilde{r}\left\{2\bar{\bv}\bar{\bv}\boldsymbol{:}(\tI+\be_r\be_r) + 4\bar{v}\bar{\bv}\bcdot\be_r\right\} + o(r) \quad \text{as} \quad \tilde{r}\searrow0. 
\end{equation}
Comparison of \eqref{remote sol steady expansion} with the general local expansion  \eqref{local behaviour} gives, for $\bv(t)\equiv \bar{\bv}$,  
\refstepcounter{equation}
$$
\label{hH steady}
\tilde{h}(t)\equiv -2\bar{v}, \quad 
\tilde{\mathbf{H}}(t)\equiv 4\bar{v}\bar{\bv}.
\eqno{(\theequation \mathrm{a},\mathrm{b}}) 
$$

One approach to obtain $\tilde{\mathbf{H}}(t)$ in the more general unsteady case is to solve the remote-region problem \eqref{remote eq}--\eqref{remote matching} numerically in three spatial dimensions plus time, starting from an initial distribution $\tilde{c}_1(\tilde{\br},0)$ that is compatible with \eqref{remote far} and \eqref{remote matching}. In this approach, $\tilde{\mathbf{H}}(t)$ is in practice determined from the instantaneous remote-region distribution $\tilde{c}_1(\tilde{\br},t)$. Given the form of the remote-region problem, it can also be conceptually considered as a time-dependent functional of the initial distribution $\tilde{c}_1(\tilde{\br},0)$ and the particle velocity $\bv(\tau)$ over the interval $0\le\tau\le t$. 

In Sec.~\ref{sec:history}, we shall develop an alternative approach by analytically solving the unsteady remote-region problem. This will lead us to an explicit representation for $\tilde{\mathbf{H}}(t)$ in terms of a `history operator', namely an integral over the history of the particle motion without explicit reference to any concentration distribution. We shall adopt this  approach in Sec.~\ref{sec:dynamics} towards studying the dynamics of force-free and forced active particles.

\subsection{Unsteady amplitude equation}
\label{ssec:amplitude}
We proceed to consider the particle-scale region, closely following the steady theory in part I. 
We begin by noting that asymptotic matching between the particle-scale concentration expansion \eqref{c expansion} taken up to order $\epsilon^2$ and the leading-order remote-region concentration approximation \eqref{remote expansion}, using the local behavior \eqref{local behaviour},  yields the following far-field conditions on the particle-scale concentration fields:
\begin{equation}
\label{c1 far}
c_1(\br,t) = \tilde{h}(t) -2\be_r\bcdot \bv(t) + o(1) \quad \text{as} \quad r\to\infty, 
\end{equation}
\begin{equation}
\label{c2 far}
c_2(\br,t) = r\left\{2\bv(t)\bv(t)\boldsymbol{:}(\tI+\be_r\be_r)+\be_r\bcdot\tilde{\mathbf{H}}(t)\right\} +o(r) \quad \text{as} \quad r\to\infty.
\end{equation}

With \eqref{c1 far}, we find from the governing equations \eqref{c bc}--\eqref{force torque balances} and \eqref{AD T} a homogeneous problem for the order-$\epsilon$ fields and particle velocities that is identical to that found in Sec.~2E of part I, except that the spatially constant term in \eqref{c1 far} is $\tilde{h}(t)$ instead of its steady-state value (\ref{hH steady}a). We thus have the general solution (cf.~(2.27)--(2.29), part I) 
\refstepcounter{equation}
\label{order 1 sol}
$$
c_1(\br,t)=\tilde{h}(t)-\bv(t)\bcdot \left(\frac{2}{r}-\frac{3}{2r^3}+\frac{1}{r^4}\right)\br 
\eqno{(\theequation \mathrm{a})}
$$
$$
\bu_1(\br,t)=-\bv(t)\bcdot \left(\tI-\frac{1}{2}\bnabla\bnabla\frac{1}{r}\right), \quad p_1(\br,t)=0,
\eqno{(\theequation \mathrm{b},\mathrm{c})}
$$
with the particle velocity $\bv(t)$ undetermined at this stage (and the corresponding angular velocity $\boldsymbol{\Omega}_1(t)=\bzero$, as it must be given our previous comment that $\boldsymbol{\Omega}(t)$ is identically zero). 
The existence of non-trivial solutions to the homogeneous order-$\epsilon$ problem is a consequence of perturbing about the threshold for instability of a force-free particle; indeed, up to the nonlinear spatially constant term $\tilde{h}(t)$, the solutions \eqref{order 1 sol} can be identified as the marginally stable modes of the full problem when linearized at the threshold $\Pen=4$ \cite{Michelin:13}. 

With \eqref{c2 far}, we find from the governing equations \eqref{c bc}--\eqref{force torque balances} and \eqref{AD T} an inhomogeneous problem for the order-$\epsilon^2$ fields and particle velocities. It is the same as that found in Sec.~4A of part I (see also Sec.~2G therein for the case of a forced particle), except for the unsteady generalization of $\tilde{\mathbf{H}}(t)$ in \eqref{c2 far} from its value (\ref{hH steady}b) for a constant particle velocity and of the force condition at this order to allow for a time-dependent force: 
\begin{equation}
\bF_2(t)=-6\pi\mathbf{f}(t).
\end{equation}

The order-$\epsilon^2$ problem is an inhomogeneous version of the homogeneous order-$\epsilon$ problem. As the latter possesses the non-trivial solutions \eqref{order 1 sol}, the former can only be solvable under certain conditions on the forcing terms. In part  I, one of us developed an explicit solvability condition for a generalized inhomogeneous problem that in the present case reduces to [cf.~part I, (3.31)]
\begin{equation}\label{solvability}
3\lim_{\lambda\to\infty}\frac{1}{\lambda^2}\oint_{r=\lambda}\mathrm{d}A\,\be_r\mathcal{R}-2\pi\mathbf{f}=\lim_{\lambda\to\infty}\oint_{1<r<\lambda}\mathrm{d}V\,\be_r\frac{\mathcal{C}}{r^2},
\end{equation}
in which 
\refstepcounter{equation}
$$
\mathcal{C}=4\bu_1\bcdot \bnabla c_1+\chi\bu_1\bcdot\bnabla\frac{1}{r}, \quad 
\mathcal{R}=2\bv\bv\boldsymbol{:}(\tI+\be_r\be_r)+\be_r\bcdot\tilde{\mathbf{H}}.
\eqno{(\theequation \mathrm{a},\mathrm{b})}
$$
The field $\mathcal{C}$ differs from that in the steady analysis of an isotropic active particle [cf.~part I, (3.32a)] only in that $\bnabla c_1$ and $\bu_1$ are here proportional to the time-dependent particle velocity $\bv(t)$ rather than necessarily a steady one (cf.~\eqref{order 1 sol}). The field $\mathcal{R}$ differs from that in the above steady analysis [cf.~part I, (3.32b)] in the unsteady generalization of $\tilde{\mathbf{H}}(t)$ from its value (\ref{hH steady}b) for a constant particle velocity. We therefore find, by either direct calculation or adapting results from part I, 
\refstepcounter{equation}
$$
\lim_{\lambda\to\infty}\oint_{1<r<\lambda}\mathrm{d}V\,\be_r\frac{\mathcal{C}}{r^2}=\chi\pi\bv,\quad 
\lim_{\lambda\to\infty}\frac{1}{\lambda^2}\oint_{r=\lambda}\mathrm{d}A\,\be_r\mathcal{R}=\frac{4\pi}{3}\tilde{\mathbf{H}}.
\eqno{(\theequation \mathrm{a},\mathrm{b})}
$$
The solvability condition \eqref{solvability} thus yields the requisite amplitude equation
\begin{equation}\label{amplitude}
\chi\bv-4\tilde{\mathbf{H}}+2\mathbf{f}=0.
\end{equation}

In its present form, the amplitude equation \eqref{amplitude} involves the vector quantity $\tilde{\mathbf{H}}(t)$. Its solution therefore entails the simultaneous solution of the remote-region problem formulated in Sec.~\ref{ssec:remote}---a linear initial-boundary-value problem in three spatial dimensions plus time. While this could be done numerically, we shall prefer to employ the analytical history-operator representation of $\tilde{\mathbf{H}}(t)$ to be developed in Sec.~\ref{sec:history} along with some of its properties. Readers not interested in this derivation may choose to skip directly to Sec.~\ref{sec:dynamics}, where we first recapitulate the amplitude equation formulated using the integral representation for $\tilde{\mathbf{H}}(t)$ and then study the dynamics of force-free and forced active particles based on that formulation. 

\section{History operator}\label{sec:history}
\subsection{Solution of the unsteady remote problem}\label{ssec:analytical}
The vector quantity $\tilde{\mathbf{H}}(t)$ appearing in the amplitude equation \eqref{amplitude} was defined in Sec.~\ref{ssec:remote} based on the local expansion \eqref{local behaviour} of the remote-region concentration field. In this section we shall develop an explicit representation for $\tilde{\mathbf{H}}(t)$ as an integral over the history of the particle motion. This representation is based on a general analytical solution to the unsteady remote-region problem \eqref{remote eq}--\eqref{remote matching}, which we derive below.

\begin{figure}[t!]
\begin{center}
\includegraphics[scale=0.5]{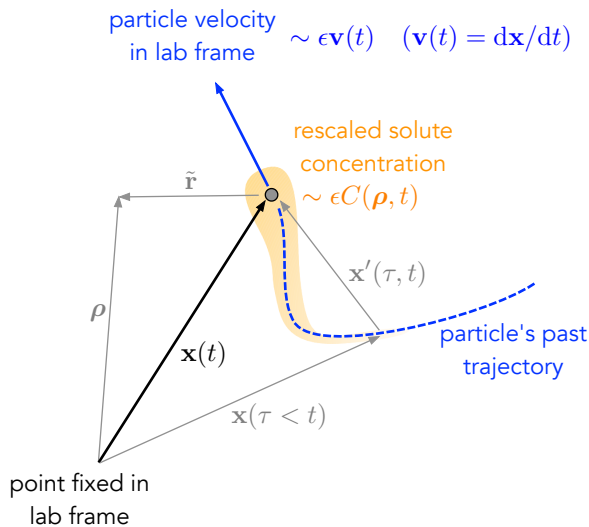}
\caption{In the lab frame, the remote-region problem consists of unsteady diffusion from a moving point source corresponding to the active particle.}
\label{fig:lab}
\end{center}
\end{figure}

It is convenient to reformulate the remote-region problem as follows. First, we combine the advection--diffusion equation \eqref{remote eq} for $\tilde{r}>0$ and the singular behavior \eqref{remote matching} to find 
\begin{equation}\label{remote eq delta} 
\pd{\tilde{c}_1}{t}-\bv \bcdot \tilde{\bnabla}\tilde{c}_1-\frac{1}{4}\tilde{\nabla}^2 \tilde{c}_1=\pi\delta(\tilde{\br}),
\end{equation}
where $\delta(\tilde{\br})$ denotes the three-dimensional Dirac delta function such that the equation holds everywhere in the sense of distributions. Second, we move from the co-moving frame to the lab frame by writing $c_1(\tilde{\br},t)=C(\boldsymbol{\rho},t)$, where $\boldsymbol{\rho}=\bx(t)+\tilde{\br}$ is a (strained) position measured from a point fixed in the lab frame, with $\bx(t)$ the corresponding particle location such that 
\begin{equation}\label{v is dot x}
\bv(t)=\frac{\mathrm{d}\bx}{\mathrm{d}t},
\end{equation}
as shown in Fig.~\ref{fig:lab}. 
The advection--diffusion equation \eqref{remote eq delta} is thus transformed into 
\begin{equation}\label{remote eq lab} 
\pd{C}{t}=\frac{1}{4}\nabla^2_{\boldsymbol{\rho}} C+\pi \delta(\boldsymbol{\rho}-\bx(t)),
\end{equation}
wherein $\bnabla_{\boldsymbol{\rho}}$ is the gradient operator with respect to $\boldsymbol{\rho}$. The lab-frame problem therefore consists of \eqref{remote eq lab} and the decay condition 
\begin{equation}\label{remote far lab}
C(\boldsymbol{\rho},t)\to0\quad \text{as} \quad |\boldsymbol{\rho}|\to\infty,
\end{equation}
which follows from \eqref{remote far}. This problem describes unsteady diffusion from a moving point source corresponding to the active particle.   

To solve the lab-frame problem we introduce the Green's function 
$G(\boldsymbol{\rho},t)$ satisfying 
\begin{equation}
\pd{G}{t}-\frac{1}{4}\nabla^2_{\boldsymbol{\rho}} G=\delta(t)\delta(\boldsymbol{\rho}),
\end{equation}
wherein $\delta(t)$ denotes the one-dimensional Dirac delta function, 
together with $G(\boldsymbol{\rho},t)=0$ for $t<0$ and $G(\boldsymbol{\rho},t)\to0$ as $\rho=|\boldsymbol{\rho}|\to\infty$. The solution is readily found using the Fourier transform and is well known to be (for $t>0$) 
\begin{equation}\label{G sol}
G(\boldsymbol{\rho},t)=\frac{1}{\pi^{3/2}t^{3/2}}e^{-\rho^2/t}.
\end{equation}
By superposition we have 
\begin{equation}\label{remote super}C(\boldsymbol{\rho},t)=\pi\int_{-\infty}^t\mathrm{d}\tau\,G(\boldsymbol{\rho}-\bx(\tau),t-\tau).
\end{equation} 
Going back to the original remote-region formulation, 
\begin{equation}\label{remote sol}
\tilde{c}_1(\tilde{\br},t)=\frac{1}{\sqrt{\pi}}\int_{-\infty}^t\mathrm{d}\tau\,\frac{1}{(t-\tau)^{3/2}}\exp\left\{-\frac{|\bx'(\tau,t)+\tilde{\br}|^2}{ t-\tau}\right\},
\end{equation}
where we define the displacement  
\begin{equation}\label{xp def}
\bx'(\tau,t)=\bx(t)-\bx(\tau)=\int_{\tau}^t\mathrm{ds}\,\bv(s),
\end{equation}
namely the position at time $t$ measured from the  position at the past time $\tau$ (see Fig.~\ref{fig:lab}).

\subsection{History operator}
We can now extract $\tilde{\mathbf{H}}(t)$ by comparing the local expansion \eqref{local behaviour} with an asymptotic expansion of \eqref{remote sol} as $\tilde{r}\searrow 0$. (This procedure will also provide $\tilde{h}(t)$ which does not enter the amplitude equation \eqref{amplitude} but the result may be useful when including some perturbation effects as well as in other problems involving unsteady convective transport, see, e.g., \cite{Feng:96,Pozrikidis:97}). The expansion for \eqref{remote sol}  is derived with the help of the subtraction 
\begin{equation}\label{remote decompose}
\tilde{c}_1(\tilde{\br},t)=\tilde{c}_1\ub{s}(\tilde{\br};\bv(t))+\Delta\tilde{c}_1(\tilde{\br},t),
\end{equation}
where the `quasi-steady' contribution $\tilde{c}_1\ub{s}(\tilde{\br},\bv(t))$ corresponds to the concentration field generated as if the particle has always been moving at its instantaneous velocity $\bv(t)$ (cf.~\eqref{remote sol steady}), while $\Delta\tilde{c}_1(\tilde{\br},t)$ is the concentration field owing to the deviation of the particle motion from that steady state. The singular small-$\tilde{r}$ behaviour of the quasi-steady contribution $\tilde{c}_1\ub{s}(\tilde{\br},\bv(t))$ is provided by \eqref{remote sol steady expansion}, with $\bv(t)$ replacing $\bar{\bv}$,
\begin{multline}\label{remote sol quasi-steady expansion}
\tilde{c}_1\ub{s}(\tilde{\br};\bv(t))= \frac{1}{\tilde{r}}-2v(t)-2{\bv(t)}\bcdot\be_r \\ +\tilde{r}\left\{2{\bv(t)}{\bv(t)}\boldsymbol{:}(\tI+\be_r\be_r) + 4v(t){\bv(t)}\bcdot\be_r\right\} + o(r) \quad \text{as} \quad \tilde{r}\searrow0. 
\end{multline}
It therefore remains to consider the small-$\tilde{r}$ behaviour of $\Delta\tilde{c}_1(\tilde{\br},t)$. To this end, we begin by noting that an integral representation of the quasi-steady contribution can be derived by substituting $\bx(t)-\bx(\tau)=(t-\tau)\bv(t)$ into the general solution \eqref{remote sol}. This gives
\begin{equation}
\tilde{c}_1\ub{s}(\tilde{\br};\bv(t))=\frac{1}{\sqrt{\pi}}\int_{-\infty}^t\frac{\mathrm{d}\tau}{(t-\tau)^{3/2}}\exp\left\{-\frac{|(t-\tau)\bv(t)+\tilde{\br}|^2}{t-\tau}\right\}.
\end{equation}
We then find from \eqref{remote sol} and the definition \eqref{remote decompose} that 
\begin{equation}\label{c diff integral}
\Delta\tilde{c}_1(\tilde{\br},t)=\frac{1}{\sqrt{\pi}}\int_{-\infty}^t\frac{\mathrm{d}\tau}{(t-\tau)^{3/2}}e^{-\frac{\tilde{r}^2}{t-\tau}}\left\{e^{-\frac{|\bx'(\tau,t)|^2}{t-\tau}}e^{-\frac{2\bx'(\tau,t)\bcdot\tilde{\br}}{t-\tau}}-e^{-(t-\tau)v^2(t)}e^{-2\bv(t)\bcdot\tilde{\br}}\right\}.
\end{equation}
Under rather mild conditions on the particle motion, discussed below, it is justified to naively expand the integrand in \eqref{c diff integral} as $\tilde{r}\searrow0$ and integrate term by term to find
\begin{equation}\label{diff expansion}
\Delta\tilde{c}_1(\tilde{\br},t)= h[\bv](t)+\tilde{\br}\bcdot \mathbf{H}[\bv](t) + o(\tilde{r}) \quad \text{as} \quad \tilde{r}\searrow0,
\end{equation}
where we define the functionals
\begin{equation}
 h[\bv](t)=\frac{1}{\sqrt{\pi}}\int_{-\infty}^t\frac{\mathrm{d}\tau}{(t-\tau)^{3/2}}\left\{e^{-\frac{|\bx'(\tau,t)|^2}{t-\tau}}-e^{-(t-\tau)v^2(t)}\right\}
\end{equation}
and
\begin{equation}\label{H def}
 \mathbf{H}[\bv](t)=\frac{2}{\sqrt{\pi}}\int_{-\infty}^t\frac{\mathrm{d}\tau}{(t-\tau)^{5/2}}\left\{(t-\tau)\bv(t)e^{-(t-\tau)v^2(t)}-\bx'(\tau,t)e^{-\frac{|\bx'(\tau,t)|^2}{t-\tau}}\right\}.
\end{equation}
Substituting the local expansions \eqref{remote sol quasi-steady expansion} and  \eqref{diff expansion} into the decomposition \eqref{remote decompose} gives the local expansion of the remote-region concentration $\tilde{c}_1(\tilde{\br},t)$ as $\tilde{r}\searrow0$ to order $\tilde{r}$. Comparing that expansion with \eqref{local behaviour} gives $\tilde{h}(t)=-2v(t)+ h[\bv](t)$ and, more importantly,  
\begin{equation}
\label{H decompose}
\tilde{\mathbf{H}}(t)=4v(t)\bv(t)+\mathbf{H}[\bv](t).
\end{equation}

We shall refer to $\mathbf{H}[\bv](t)$ as the history operator, with the notation $[\bv](t)$ indicating a functional dependence upon $\bv(\tau)$ over past times $\tau\le t$. In equations where the dependence on $t$ is implicit, we shall still indicate the functional dependence by writing $\mathbf{H}[\bv]$. The history operator clearly vanishes for steady rectilinear motion. Hence, \eqref{H decompose} decomposes the vector quantity $\tilde{\mathbf{H}}(t)$ into quasi-steady and inherently unsteady parts. 

A subtle point is that the definition of $\tilde{\mathbf{H}}(t)$ via the remote-region initial-boundary-value problem (cf.~Sec.~\ref{ssec:remote}) is slightly more general than that based on the decomposition \eqref{H decompose} and the history operator \eqref{H def}. Indeed, the singular integral in \eqref{H def} converges only if $\bx'(\tau,t)=\bv(t)(t-\tau)+\mathcal{O}((t-\tau)^{q})$ as $\tau\nearrow t$, with $q>3/2$. (This condition is actually necessary for justifying the straightforward expansion of \eqref{c diff integral} leading to \eqref{diff expansion}.)  
With reference to \eqref{xp def}, this condition is clearly satisfied for any smooth particle trajectory, in which case $q\ge2$. We will also consider scenarios involving a force discontinuity, say at time $t=t_0$, in which case we shall find that $\bv(t)$ is continuous but varies like $(t-t_0)^{1/2}$ as $t\searrow t_0$. The above condition on $\bx'(\tau,t)$  still holds as the left derivative of $\bv(t)$ exists at any fixed $t$ including $t_0$. 

\subsection{Slowly varying velocity}
\label{ssec:slowly}
If $\bv(t)$ varies slowly enough, the history operator can sometimes be expanded in terms of differential operators, as follows. The first term in the integrand in \eqref{H def} attenuates exponentially over the recent history $t-\tau=\mathcal{O}(1/v^2(t))$. If $\bv(t)$ is non-vanishing and varies on a time scale $T\gg1/v^2$, then considering the `recent history' $t-\tau=\mathcal{O}(1/v^2(t))$, we can expand
\begin{equation}\label{xp expansion}
\bx'(\tau,t)=\int_{\tau}^t \mathrm{d}s\,\bv(s) = (t-\tau)\bv-\frac{1}{2}(t-\tau)^2\dot{\bv}+\frac{1}{6}(t-\tau)^3\ddot{\bv}+\cdots, 
\end{equation}
with each subsequent term being $1/(v^2T)$ smaller in order of magnitude. (An upper dot denotes differentiation with respect to $t$, with the velocity $\bv$ and its derivatives evaluated at time $t$ unless stated otherwise.) Furthermore, in the same interval we have the expansion
\begin{multline}\label{xp exp expansion}
e^{-\frac{|\bx'(\tau,t)|^2}{t-\tau}}=e^{-(t-\tau)v^2}\left\{1+(t-\tau)^2\dot{\bv}\bcdot\bv+\left[\frac{1}{2}(t-\tau)^4(\dot{\bv}\bcdot\bv)^2
\right. \right.\\\left.\left.
-(t-\tau)^3\left(\frac{\dot{v}^2}{4}+\frac{\ddot{\bv}\bcdot\bv}{3}\right) \right]+\cdots\right\}.
\end{multline}
It is therefore clear that the second term in the integrand in \eqref{H def} attenuates exponentially on the same time scale as the first, whereby the contribution of the recent history is local. Thus, substituting \eqref{xp expansion} and \eqref{xp exp expansion} into \eqref{H def} and integrating term by term yields
\begin{equation}\label{slow expansion}
\mathbf{H}[\bv]=\frac{\tI-\hat{\bv}\hat{\bv}}{v}\bcdot \dot{\bv}+\left[3\frac{|\dot{\bv}|^2\hat{\bv}-5(\hat{\bv}\bcdot\dot{\bv})^2\hat{\bv}+2(\hat{\bv}\bcdot\dot{\bv})\dot{\bv}}{8 v^4}-\frac{(\tI-3\hat{\bv}\hat{\bv})\bcdot\ddot{\bv}}{6v^3}\right]
+\cdots,
\end{equation}
wherein $\hat{\bv}=\bv/v$. 

Besides the condition that $\bv(t)$ varies sufficiently slowly, the expansion \eqref{slow expansion} can fail in situations where the particle returns to a location it has already visited. This is because $\bx'(\tau,t)$ is small near those past moments and hence there can be localized contributions to the integral in \eqref{H def} from the non-recent history that are only algebraically small. 

\subsection{Linearization}\label{ssec:linear}
We next consider the linearization of the history operator about either the stationary state or steady rectilinear motion. Writing $\bv(t)=\bv_0+\delta\bv(t)$, wherein $\bv_0$ is a constant vector (possibly the zero vector) of magnitude $v_0$ and direction $\hat{\bv}_0$, and $\delta\bv(t)$ a perturbation, we find from \eqref{H def} that, to linear order in the perturbation,
\begin{equation}\label{dH def}
\mathbf{H}[\bv]=\delta\mathbf{H}[\delta\bv;\bv_0],
\end{equation}
where  
\begin{equation}\label{dH basic}
\delta\mathbf{H}[\delta\bv;\bv_0]=
\frac{2}{\sqrt{\pi}}\int_{-\infty}^t\mathrm{d}\tau\,\frac{e^{-(t-\tau)v_0^2}}{(t-\tau)^{5/2}}\left\{\tI-2(t-\tau)\bv_0\bv_0\right\}\bcdot\int_{\tau}^t\mathrm{d}s\,\left\{\delta\bv(t)-\delta\bv(s)\right\}.
\end{equation}

For linearization about the stationary state, we set $\bv_0=\bzero$. In that case, interchanging the order of integration in \eqref{dH basic} and performing the integration in $\tau$ gives
\begin{equation}\label{H linear zero}
\delta\mathbf{H}[\delta\bv;\bzero]=\frac{4}{3\sqrt{\pi}}\int_{-\infty}^t\mathrm{d}s\,\frac{\delta\bv(t)-\delta\bv(s)}{(t-s)^{3/2}}.
\end{equation}
We note that this linearization fails for a perturbation that decayed too fast in the far past, since with the integrand in \eqref{H linear zero} behaving like $\delta\bv(s)/|s|^{3/2}$ as $s\to-\infty$ the contribution of the far past could diverge. In particular, this linearization can describe exponentially growing but not exponentially decaying perturbations. Thus, assuming an exponential perturbation of the form $\delta\bv(t)=\mathbf{A}e^{\sigma t}$ 
wherein $\mathbf{A}$ is a constant vector and $\sigma$ a complex growth rate, \eqref{H linear zero} gives 
\begin{equation}\label{dH exp stat}
\delta\mathbf{H}[\mathbf{A}e^{\sigma t};\bzero]=\frac{8}{3}\sigma^{1/2}\mathbf{A}e^{\sigma t}
\end{equation}
for $\mathop{\mathrm{Re}}\sigma\ge0$, 
where the square root takes its principal value. 

For linearization about a steady rectilinear motion, we have $v_0\ne0$. In that case, interchanging the order of integration in \eqref{dH basic} and performing the integration in $\tau$ gives, after making the change of variables $q=v_0^2(t-s)$,
\begin{equation}\label{dH def steady motion}
\delta\mathbf{H}[\delta\bv;\bv_0]=  \frac{2v_0}{\sqrt{\pi}}\int_0^{\infty}\mathrm{d}q\,\left\{\tI f_1(q)-2\hat{\bv}_0\hat{\bv}_0f_2(q)\right\}\bcdot\left\{\delta\bv(t)-\delta\bv(t-q/v_0^2)\right\},
\end{equation}
in which  
\refstepcounter{equation}
$$
\label{fs def}
f_1(q)=\frac{2(1-2q)e^{-q}}{3q^{3/2}}+\frac{4\sqrt{\pi}}{3}\,\mathrm{erfc}(q^{1/2}), \quad f_2(q)=2q^{-1/2}e^{-q}-2\sqrt{\pi}\,\mathrm{erfc}(q^{1/2}),
\eqno{(\theequation \mathrm{a},\mathrm{b})}
$$
with $\mathrm{erfc}$ the complementary error function \cite{Abramowitz:book}. As $q\to0$, $f_1(q)$ and $f_2(q)$ scale like $q^{-3/2}$ and $q^{-1/2}$, respectively; this confirms that the integration in \eqref{dH def steady motion} converges as $q\to0$, as the second curly bracket on the right-hand side of \eqref{fs def} scales like $q$ in that limt.  As $q\to\infty$, these functions scale like $q^{-5/2}e^{-q}$ and $q^{-3/2}e^{-q}$. Hence, the linearization about a steady rectilinear motion fails for perturbations that decay at a too high exponential rate in the far past. Assuming an exponential perturbation as before, we find from \eqref{dH def steady motion} 
\begin{equation}\label{dH exp rect}
\delta\mathbf{H}[\mathbf{A}e^{\sigma t};\bv_0]=\frac{8v_0}{3\sigma'}\left\{\tI \left[(1+\sigma')^{3/2}-1-\frac{3}{2}\sigma'\right]-\frac{3}{2}\hat{\bv}_0\hat{\bv}_0\left[(1+\sigma')^{1/2}-1\right]^2\right\}\bcdot \mathbf{A}e^{\sigma t}, 
\end{equation} 
for $\mathop{\mathrm{Re}}\sigma'\ge-1$, wherein $\sigma'=\sigma/v_0^2$. 
We note that \eqref{dH exp rect}  degenerates to \eqref{dH exp stat} as $v_0\to0$ and agrees with a linearization of the slowly varying approximation \eqref{xp exp expansion} in the limit $\sigma\to0$.  

\section{History-dependent dynamics}\label{sec:dynamics}
\subsection{Weakly nonlinear model and numerical method}\label{ssec:recap}
Let us recapitulate the weakly nonlinear model we have developed for the slow dynamics of a force-free or weakly forced isotropic active particle near the threshold for spontaneous motion. Recall the definition $\Pen=4+\epsilon \chi$, with $0<\epsilon\ll1$ and $\chi$ a rescaled bifurcation parameter, and that the slow time $t$ is normalized by $a_*/(\epsilon^2u_*)$. The particle velocity $\bv(t)$ in the lab frame (normalized by $\epsilon u_*$) satisfies the unsteady amplitude equation (cf.~\eqref{amplitude} and \eqref{H decompose}) gives
\begin{equation}\label{amplitude repeat}
(\chi-16v)\bv-4\mathbf{H}[\bv]+2\mathbf{f}=\bzero,
\end{equation}
where $v(t)$ is the magnitude of $\bv(t)$, $\mathbf{f}(t)$ is the generally time-dependent force on the particle (normalized by $6\pi\epsilon^2\eta_*u_*a_*$) and the history operator is given by
\begin{equation}\label{H repeat}
\mathbf{H}[\bv](t)=\frac{2}{\sqrt{\pi}}\int_{-\infty}^t\frac{\mathrm{d}\tau}{(t-\tau)^{5/2}}\left\{(t-\tau)\bv(t)e^{{-(t-\tau)v^2(t)}}-\bx'(\tau,t)e^{-\tfrac{|\bx'(\tau,t)|^2}{t-\tau}}\right\},
\end{equation}
in which 
\begin{equation}\label{xp repeat}
\bx'(\tau,t)=\bx(t)-\bx(\tau)=\int_{\tau}^t\mathrm{ds}\,\bv(s)
\end{equation}
is the position of the particle (normalized by $a_*/\epsilon$) at time $t$ measured from the position at the past time $\tau$, with $\bx(t)$ being the position of the particle measured from a fixed point in the lab frame (see Fig.~\ref{fig:lab}). The history operator $\mathbf{H}[\bv](t)$ captures the relaxation of the particle's concentration wake, vanishing for a constant particle velocity in which case the wake is steady in a co-moving frame. 

We call attention to the fact that only the product $\epsilon\chi$, in which $\epsilon$ is constrained to be positive, has physical significance, rather than $\epsilon$ or $\chi$ separately. We could generally choose $\chi=\pm1$, in which case $\epsilon$ is the absolute magnitude of $\Pen-4$ and the magnitude of the force is characterized by that of $\mathbf{f}(t)$. A different viewpoint is that $\epsilon^2$ characterizes the magnitude of the force, in which case  $\chi$ is a free parameter. 

For the remainder of this section we study the above weakly nonlinear model both analytically and numerically. In Sec.~\ref{ssec:steady}, we review its steady-state solutions, already known from part I and earlier works \cite{Morozov:19,Morozov:19b,Saha:21}. In Sec.~\ref{ssec:stability} we analyze the stability of those steady states by employing the linearization of the history operator developed in Sec.~\ref{ssec:linear}, and illustrate the results via numerical simulations. In Sec.~\ref{ssec:alignment}, we study the alignment dynamics of a spontaneously moving particle following the sudden application of an external force, supplementing numerical simulations with  analytical approximations based on the expansion developed in Sec.~\ref{ssec:slowly} of the history operator for a slowly varying particle velocity.

For the numerical simulations, we assume that $\bv(t)$ is known and constant for $t<0$, corresponding to a steady state of \eqref{amplitude repeat} for some constant forcing. The main step in the numerical scheme takes a sequence of values $\bx(t_i)$ at (adaptable) times $t_0=0,t_1,\ldots,t_{n-1}$ and calculates the value $\bx(t_n)$ using Newton iteration. The history operator is evaluated using the analytical expressions derived in appendix \ref{app_s:tail} for the prescribed `tail' $\tau<0$ and the trapezoidal rule for $0\le\tau<t$. For the latter integration, we use either the basic form \eqref{H repeat} of the history operator where the integrand exhibits an integrable $1/(t-\tau)^{1/2}$ singularity as $\tau\nearrow t$, which requires estimating the velocity $\bv(t_n)$, or a regularized version of the history operator developed in appendix \ref{app:reg}, which requires estimating also the acceleration $\ba(t_n)=(\mathrm{d}\bv/\mathrm{d}t)(t_n)$; the regularization improves the order of the scheme from $\mathcal{O}(\Delta t^{1/2})$ to $\mathcal{O}(\Delta t^{3/2})$, wherein $\Delta t$ is the time step. We generally assume that $\mathbf{f}(t)$ is smooth but the code does allow for discontinuity at $t=0$; in that case, we employ the early time asymptotic expansion derived in appendix \ref{app_s:early} to determine the values of $\bx(t_i)$ for the first few initial times.  

 \subsection{Steady-state solutions}\label{ssec:steady}
We look for steady state solutions $\bv(t)=\bar{\bv}$ of the amplitude equation \eqref{amplitude repeat}, of magnitude $\bar{v}$ and direction $\hat{\bar{\bv}}$. For this purpose, we assume a constant force $\mathbf{f}(t)={f}\unit$, with $f\ge0$ and $\unit$ a unit vector. As the history operator $\mathbf{H}[\bv](t)$ vanishes for a constant velocity, \eqref{amplitude repeat} reduces to the steady amplitude equation 
\begin{equation}\label{steady amplitude}
(\chi-16\bar{v})\bar{\bv}+2{f}\unit=\bzero,
\end{equation}
which was already derived and studied in part I. It was also derived in \cite{Saha:21} with the \textit{a priori} assumption that the motion is collinear with the force, and earlier in \cite{Rednikov:94,Rednikov:94b} for a closely related active-drop model. 

Consider first the case $f=0$ corresponding to a freely suspended particle. For all $\chi$, we find the trivial steady state $\bar{\bv}=\bzero$. Additionally, for $\chi>0$ we find non-trivial steady states corresponding to spontaneous rectilinear motion with speed $\bar{v}=\chi/16$ and arbitrary direction. 

Consider next the case $f>0$ corresponding to a constant external force. In contrast to the force-free case, where the direction of steady spontaneous motion is arbitrary, it is evident from \eqref{steady amplitude} that in the forced case steady motion is necessarily collinear with the force no matter the force magnitude. We accordingly write $\bar{\bv}=\bar{v}_{\parallel}\unit$, with the component $\bar{v}_{\parallel}$ found to satisfy the bifurcation relation
\begin{equation}\label{1d bifurcation}
\chi\bar{v}_{\parallel}-16|\bar{v}_{\parallel}|\bar{v}_{\parallel}+2{f}=0.
\end{equation}
For all $\chi$, we find the parallel $(\bar{v}_{\parallel}>0)$ branch
\begin{equation}\label{parallel}
v_{\parallel} = \frac{1}{32}\left(\chi+\sqrt{\chi^2+128{f}}\right).
\end{equation}
For $\chi>8\sqrt{2{f}}$, we additionally find the two anti-parallel  $(\bar{v}_{\parallel}<0)$ branches 
\begin{equation}\label{aparallel}
v_{\parallel} = -\frac{1}{32}\left(\chi\pm \sqrt{\chi^2-128{f}}\right),
\end{equation}
which are degenerate at $\chi=8\sqrt{2f}$. 

\begin{figure}[t!]
\begin{center}
\includegraphics[scale=0.4]{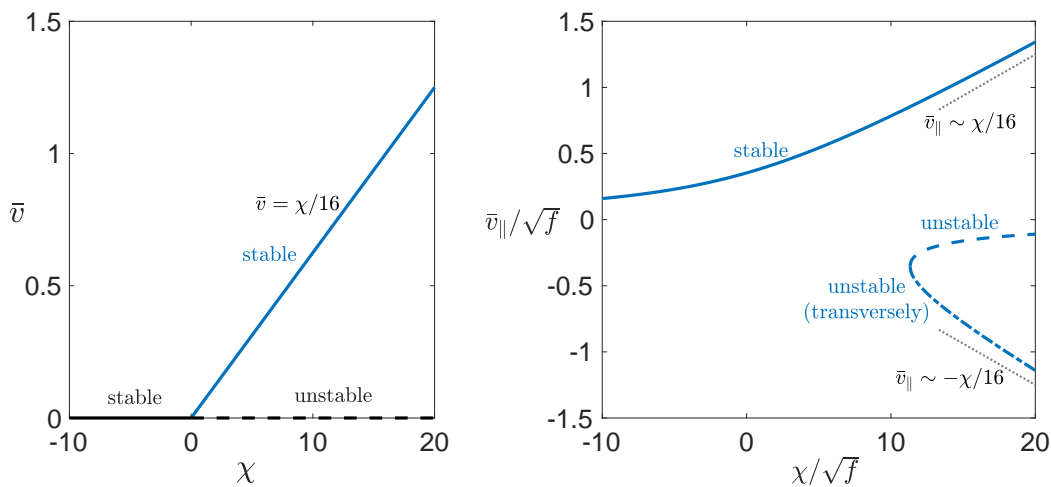}
\caption{Steady bifurcation diagrams for a freely suspended (a) and forced (b) active particle, as found in Sec.~\ref{ssec:steady}. The indicated stability characteristics are deduced in Sec.~\ref{ssec:stability}.}
\label{fig:bif}
\end{center}
\end{figure}

Steady bifurcation diagrams in the free and forced cases are depicted in Fig.~\ref{fig:bif}. Clearly in the latter case the dependence on $f$ can be scaled out by noting that $\bar{v}_{\parallel}/\sqrt{f}$ is a function of $\chi/\sqrt{f}$. 

\subsection{Linear stability}\label{ssec:stability}
We next consider the linear stability of the steady states found above. Upon writing $\bv(t)=\bar{\bv}+\delta\bv(t)$, with $\bar{\bv}$ a solution of the steady amplitude equation \eqref{steady amplitude} and $\delta\bv(t)$ an infinitesimal perturbation, a linearization of the unsteady amplitude equation \eqref{amplitude repeat} 
\begin{equation}\label{amplitude linear}
\left\{\chi-16\bar{v}\left(\tI+\hat{\bar{\bv}}\hat{\bar{\bv}}\right)\right\}\bcdot\delta\bv
-4\,\delta\mathbf{H}[\delta\bv;\bar{\bv}]=\bzero,
\end{equation}
where the linearized history operator $\delta\mathbf{H}[\delta\bv;\bar{\bv}]$ is provided by \eqref{H linear zero} for $\bar{\bv}=\bzero$ and \eqref{dH def steady motion} for $\bar{\bv}\ne\bzero$. Specifically, we look for exponential solutions of the form $\delta\bv(t)=\mathbf{A} e^{\sigma t}$, with $\mathbf{A}$ a constant vector and $\sigma$ a complex-valued growth rate. We recall from Sec.~\ref{ssec:linear} that the linearization of the history operator is only valid for perturbations that don't decay too quickly; in particular, for exponential solutions of the above form, $\delta\mathbf{H}[\delta\bv;\bar{\bv}]$ reduces to \eqref{dH exp stat} for $\bar{v}=0$, requiring $\mathop{\mathrm{Re}}\sigma\ge0$, and \eqref{dH exp rect} for $\bar{v}\ne0$, requiring $\mathop{\mathrm{Re}}\sigma\ge-\bar{v}^2$. Existence of exponentially growing solutions implies instability, of course. We shall also assume that the absence of either constant or exponentially growing solutions implies stability (if a constant solution is the fastest growing solution then we shall say that the stability is marginal); we discuss this assumption below in Sec.~\ref{sssec:stabilitydiscussion}. 

\subsubsection{Stability of the stationary state in the force-free scenario} 
Consider first the stationary state that exists for all $\chi$ in the unforced case. Substituting \eqref{dH exp stat} into \eqref{amplitude linear} gives exponential perturbations with arbitrary amplitude $\mathbf{A}$ and growth rate $\sigma$ satisfying 
\begin{equation}\label{sigma stat}
\sigma^{1/2}=\frac{3\chi}{32}
\end{equation}
and $\mathop{\mathrm{Re}}\sigma\ge0$, with the square root taking its principal value. We accordingly conclude that the stationary state is unstable for $\chi>0$, marginally stable for $\chi=0$ and stable for $\chi<0$. 

\subsubsection{Stability of steady rectilinear-motion states}
We next derive general results for $\bar{\bv}\ne\bzero$ which we will subsequently apply to infer the stability of the steady spontaneous-motion states in the force-free and forced scenarios. By substituting \eqref{dH exp rect} into \eqref{amplitude linear}, we find the condition 
\begin{equation}\label{sigma prime eq}
\frac{(1+\sigma')^{3/2}-1}{\sigma'}+3\delta_{\parallel}\frac{(1+\sigma')^{1/2}-1}{\sigma'}=\chi',
\end{equation}
in which 
\refstepcounter{equation}
$$
\label{sigma chi prime def}
\sigma'=\sigma/\bar{v}^2, \quad  \chi'=3\chi/(32\bar{v}),
\eqno{(\theequation \mathrm{a},\mathrm{b})}
$$
 and $\delta_{\parallel}$ equals $1$ for longitudinal perturbations $\mathbf{A}\parallel \bar{\bv}$ and vanishes for transverse perturbations $\mathbf{A}\perp\bar{\bv}$. Recall that in the present case the growth rate must also satisfy $\mathop{\mathrm{Re}}\sigma'\ge-1$. 

For longitudinal perturbations, we find that \eqref{sigma prime eq} has no solutions for $\chi'<1$ and two solutions for $\chi'>1$, degenerate at $\chi'=1$, given by
\begin{equation}\label{sigma prime L two solutions}
{\sigma_L'}^{\!\pm}=\frac{1}{2}\left\{\chi'^2-9\pm (\chi'-1)(\chi'-3)^{1/2}(\chi'+5)^{1/2}\right\}.
\end{equation}
For $1<\chi'<3$, the solutions \eqref{sigma prime L two solutions} are complex conjugate with a negative real part (with $\mathop{\mathrm{Re}}\sigma'\ge-1$ only at and beyond an intermediate $\chi'$). For $\chi'=3$, they degenerate to the zero solution. For $\chi'>3$, there is one positive solution and one negative solution (both with $\mathop{\mathrm{Re}}\sigma'\ge-1$). We conclude that the spontaneous-motion state is longitudinally unstable for $\chi'>3$, marginally longitudinally stable for $\chi'=3$ and longitudinally stable for $\chi'<3$. 

For transverse perturbations, we find that \eqref{sigma prime eq} has no solutions for $\chi'<1$ and just one (real) solution for $\chi'\ge1$ given by
\begin{equation}\label{sigma prime T}
\sigma'_T=\frac{1}{2}\left\{\chi'^2-3+(\chi'+3)^{1/2}(\chi'-1)^{3/2} \right\}.
\end{equation}
This solution is negative for $1\le \chi'<3/2$ (between $-1$ and $0$, hence satisfying the condition $\mathop{\mathrm{Re}}\sigma'\ge-1$), vanishes for $\chi'=3/2$, and positive for $\chi'>3/2$. Thus, the spontaneous-motion state is transversely unstable for $\chi'>3/2$, marginally transversely stable for $\chi'=3/2$ and transversely stable for $\chi'<3/2$. 

\subsubsection{Stability of spontaneous motion in the force-free scenario}
We next apply the above results to the non-trivial steady solutions of \eqref{steady amplitude}. In the unforced scenario, substituting $\bar{v}=\chi/16$ (wherein $\chi>0$) into (\ref{sigma chi prime def}b) gives $\chi'=3/2$ (cf.~(\ref{sigma chi prime def}b)). We accordingly conclude that the steady spontaneous-motion states in the force-free case are longitudinally stable and transversely marginally stable. Clearly an infinitesimal transverse perturbation simply leaves the particle velocity in the isotropic manifold of steady states. 

\subsubsection{Stability of rectilinear motion in the forced scenario}
Consider next the forced case. For the parallel state, which exists for all $\chi$ and is given by \eqref{parallel}, (\ref{sigma chi prime def}b) gives $\chi'=\chi_P'(\chi/\sqrt{f})$, where
\begin{equation}\label{chi prime p}
\chi_P'(x)=\frac{3x}{x+\sqrt{x^2+128}}.
\end{equation}
Since $\chi'_P<3/2$, we conclude that the parallel state is stable. For the anti-parallel states, which exist for $\chi>8\sqrt{2f}$ and are provided by \eqref{aparallel}, (\ref{sigma chi prime def}b) gives $\chi'=\chi_{AP,1}'(\chi/\sqrt{f})$ for the larger-magnitude branch and $\chi'=\chi_{AP,2}'(\chi/\sqrt{f})$ for the smaller-magnitude branch, where 
\refstepcounter{equation}
$$
\label{chi prime ap}
\chi'_{AP,1}(x)=\frac{3x}{x+\sqrt{x^2-128}}, \quad \chi'_{AP,2}(x)=\frac{3x}{x-\sqrt{x^2-128}}.
\eqno{(\theequation \mathrm{a},\mathrm{b},\mathrm{c})}
$$
Since $3/2<\chi'_{AP,1}<3$ the larger-magnitude anti-parallel state is longitudinally stable but transversely unstable, so overall unstable. Furthermore, since $\chi'_{AP,2}>3$ the smaller-magnitude anti-parallel is unstable (both longitudinally and transversely). 

\subsubsection{Discussion and numerical demonstrations}
\label{sssec:stabilitydiscussion}
\begin{figure}[t!]
\begin{center}
\includegraphics[scale=0.5]{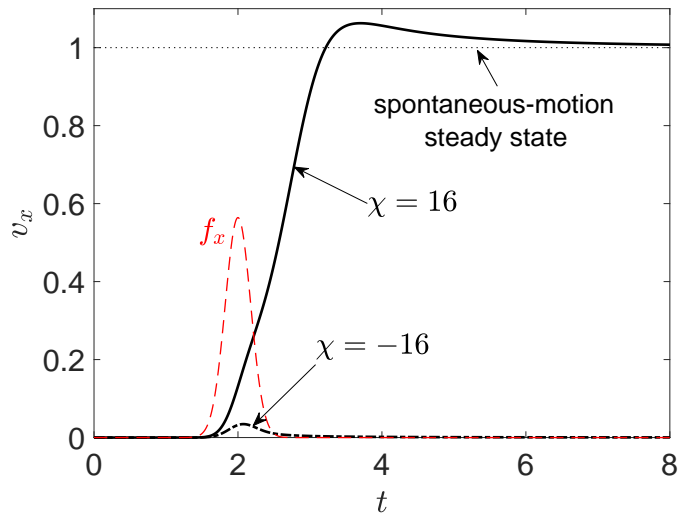}
\caption{The dynamical response of an isotropic active particle, initially at rest, to a uni-directional gaussian force input (chosen arbitrarily in the $\be_x$ direction). Depending on the sign of $\chi$, the particle is observed to either relax back to its stationary state or to approach a state of steady spontaneous motion in the direction of the force disturbance.}
\label{fig:stab_stat}
\end{center}
\end{figure}

\begin{figure}[h!]
\begin{center}
\includegraphics[scale=0.5]{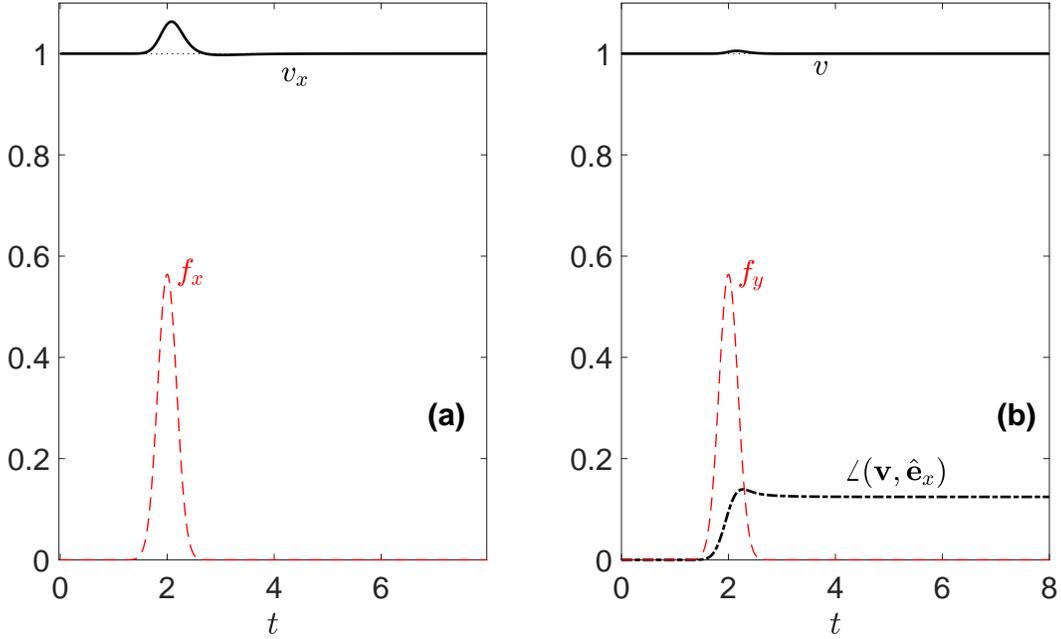}
\caption{The dynamical response of an isotropic active particle at $\chi=16$, initially at the steady spontaneous-motion state $\bar{\bv}=\be_x$, to a uni-directional gaussian force input either in the longitudinal $\be_x$ direction (a) or a transverse direction (b), the latter chosen arbitrarily as the $\be_y$ direction. In (a) the particle velocity $\bv(t)=v_x(t)\be_x$ relaxes back to its initial spontaneous-motion state. In (b) the particle speed $v$ relaxes back to its initial value but the direction of motion at long times differs from the initial one, namely the particle settles into spontaneous motion in a different direction.}
\label{fig:stab_spont}
\end{center}
\end{figure}

\begin{figure}[h!]
\begin{center}
\includegraphics[scale=0.5]{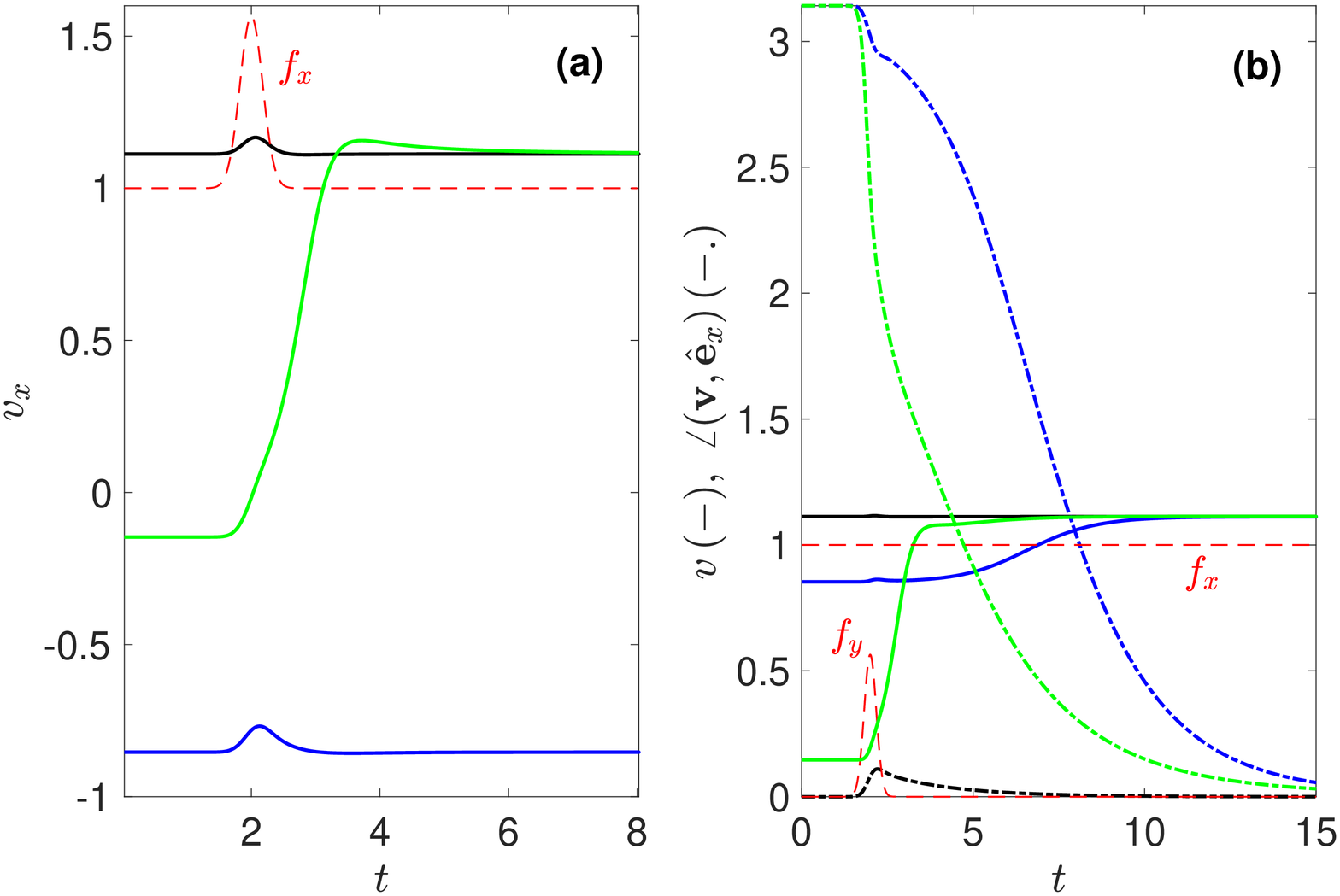}
\caption{An isotropic active particle at $\chi=16$ is initially at a steady state $\bar{\bv}=\bar{v}_{\parallel}\be_x$ corresponding to a constant force $\mathbf{f}=\be_x$ (black: parallel state, blue: larger-magnitude anti-parallel state, green: smaller-magnitude anti-parallel state). We show the dynamical response to a uni-directional gaussian force-disturbance input either in the longitudinal $\be_x$ direction (a) or a transverse direction (b), the latter chosen arbitrarily as the $\be_y$ direction. In (a) the particle velocity $\bv(t)=v_x(t)\be_x$  relaxes back to its initial state for the parallel and larger-magnitude anti-parallel states, or approaches the parallel state for the smaller-magnitude anti-parallel state. In (b) the particle velocity (speed and direction) approaches the parallel state in all cases.}
\label{fig:stab_force}
\end{center}
\end{figure}

Let us summarize the above stability results, which we have indicated in the steady bifurcation diagrams in Fig.~\ref{fig:bif}. In the unforced case, the stationary state is unstable for $\chi>0$, while the spontaneous-motion states are longitudinally stable and transversely marginally stable (as a transverse perturbation leaves the particle velocity in the isotropic manifold of spontaneous-motion states). In the forced case, the parallel state is stable; while both anti-parallel states are unstable, the larger-magnitude anti-parallel state is longitudinally stable. 

To test our stability predictions, in Figs.~\ref{fig:stab_stat}--\ref{fig:stab_force} we numerically simulate particles that are disturbed from steady-state motion (force-free or forced) by finite localized (Gaussian) force perturbations. We find that our predictions qualitatively agree with the observed dynamics in all cases. 

It is useful to compare our stability predictions with related results in the literature. The main stability result in the literature is the linear-stability analysis of Michelin \textit{et al}.~of the stationary state \cite{Michelin:13}, starting from the full model as formulated in Sec.~\ref{sec:formulation} in the case where there is no force. (A generalization of this stability analysis to a closely related model of an active drop is given in \cite{Morozov:19}.) They look for exponential solutions and find that all growth-rate eigenvalues are real. In particular, they find that exponential solutions with a positive growth rate exist for $\Pen>4$, in agreement with our finding that the stationary state is unstable for $\chi>0$. Moreover, we find that their implicit solution for positive growth rates (Eq.~(14) therein) degenerates to our solution \eqref{sigma stat} as $\Pen\searrow 4$. As further discussed below, they also find that for all P\'eclet numbers any negative growth rate is possible. 

We are not aware of a comparable linear-stability analysis in the literature for the rectilinear-motion states in either the force-free or forced cases. In the force-free case, it is natural to simply assume that the spontaneous-motion states are stable (at least in some interval of P\'eclet numbers above the threshold). We note that in appendix B of \cite{Morozov:19b} the authors claim to confirm the linear stability of the spontaneous-motion state for a closely related model of an active drop, but their calculation appears to rely on an axisymmetric weakly nonlinear analysis that incorrectly neglects the history effect. 

In the forced case, it has been incorrectly guessed in \cite{Saha:21} that both the parallel and larger-magnitude anti-parallel states are stable, while the smaller-magnitude anti-parallel state is unstable (we have shown that the larger-magnitude anti-parallel state is transversely unstable). This claim has been referenced in a recent review on active drops \cite{Michelin:22} and in a recent axisymmetric numerical study \cite{Kailasham:22}. We note that earlier weakly nonlinear analyses of forced active drops have concluded similarly to here that there is a stable parallel state and, in some cases, also two unstable anti-parallel states.  In \cite{Rednikov:94,Rednikov:94b}, where the active-drop model is closely related to our particle formulation, only a heuristic relative-stability argument is given which is strictly based on calculating steady states. In \cite{Rednikov:95}, this conclusion is reached based on an unsteady weakly nonlinear analysis, but for an active-drop model that does not involve bulk transport and hence differs fundamentally from our particle model in that there is no history effect. 

Finally, we discuss the limitations of our linear stability analysis, within the framework of the weakly nonlinear model. Firstly, we have assumed above that if the growth-rate equation \eqref{sigma stat} or \eqref{sigma prime eq} does not have a solution with positive real part then all perturbations decay, but this does not reveal the rate of decay in the cases where there is no solution at all. To analyse the general evolution of linear perturbations with time, we can consider the response of the particle to a force impulse (i.e.~similar to the numerical demonstrations in Figs.~\ref{fig:stab_stat}--\ref{fig:stab_force} but with infinitesimal localized perturbations). In Appendix \ref{app:impulse}, we perform this analysis using Laplace transforms and contour deformation, and show that if the equations do not have a solution satisfying the necessary condition $\sigma \geq -\bar{v}^2$, then the linear perturbations decay like $e^{-\bar{v}^2 t}$ times a power of $t$ as $t \to \infty$.

Secondly, the formulation with the history operator \eqref{H decompose} assumes that the perturbations to the concentration field are solely generated by perturbations to the particle motion. In order to consider more general perturbations to the concentration field, one would have to employ the amplitude equation in the form \eqref{amplitude}, with $\tilde{\mathbf{H}}(t)$ defined by the unsteady remote-region concentration problem of Sec.~\ref{ssec:remote} rather than the history operator via \eqref{H decompose}. While we do not present this analysis here, we note that using this approach to seek perturbations with time dependence $e^{\sigma t}$ to the stationary state in the force-free case recovers the result \eqref{sigma prime eq} for $\chi > 0$, while for $\chi < 0$ any negative growth rate is possible, by an appropriate choice of the initial concentration distribution, which is in agreement with the findings in \cite{Michelin:13}.

\subsection{Alignment dynamics}\label{ssec:alignment}
Consider next an isotropic active particle subjected to the external force $\mathbf{f}(t)=f\mathcal{H}(t)\unit$, where $f>0$ is the force magnitude, $\mathcal{H}(t)$ is the unit-step function and $\unit$ is a fixed unit vector. For $t<0$, the particle can be assumed to be either stationary or, for $\chi>0$, in a steady spontaneous state moving with speed $\chi/16$ in some direction. The motion is clearly confined to the plane spanned by the force and the initial spontaneous motion; if the particle is initially stationary, then the motion is collinear with the force. 

The stability analysis above suggests that typically the particle will approach with time the parallel steady state for a constant force (cf.~\eqref{parallel}), which is stable and exists for all $\chi$; if the particle is initially stationary, or if its initial spontaneous motion happens to be collinear with the force, then depending on the initial state and if $\chi>8\sqrt{2f}$ the terminal state can also be the larger-magnitude anti-parallel state (cf.~\eqref{aparallel}), which is longitudinally stable but transversely unstable. 

\begin{figure}[t!]
\begin{center}
\includegraphics[scale=0.5]{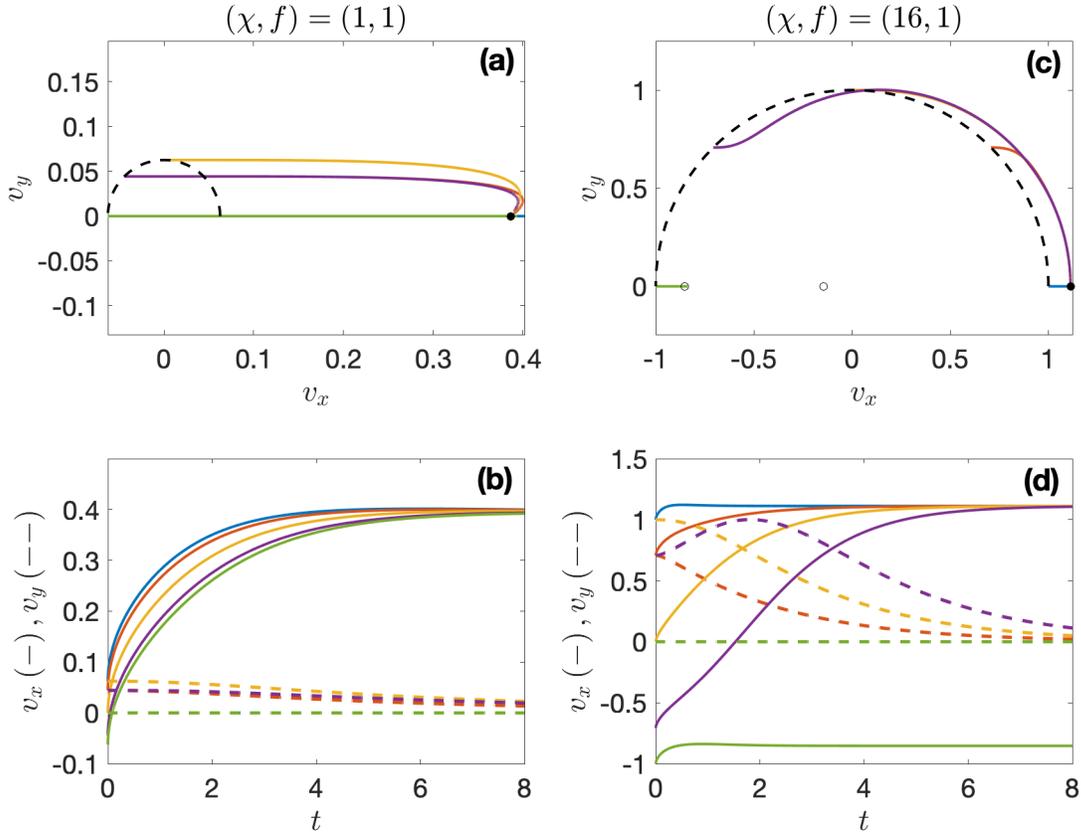}
\caption{Dynamical alignment of an isotropic active particle initially at a steady-spontaneous state $\bar{\bv}=(\chi/16)\hat{\bp}$ with some direction $\hat{\bp}$, subjected to a unit-step force in the $\be_x$ direction. We take $\chi=1$ (a,b) or $\chi=16$ (c,d) and choose several $\hat{\bp}$ in the $x\,$--$\,y$ plane (distinguished by color) such that $\bv(t)=v_x(t)\be_x+v_y(t)\be_y$. In (a,c) we plot the path in the $v_x\,$--$\,v_y$ plane; the black dashed semi-circles correspond to the unforced spontaneous-motion steady states, while the solid and hollow circular symbols respectively correspond to stable and unstable steady states with the force on. In (b,d) we plot the velocity components as a function of time $t$.}
\label{fig:alignment}
\end{center}
\end{figure}

We demonstrate the alignment dynamics described above in Fig.~\ref{fig:alignment} by running a series of simulations starting from initial spontaneous-motion states in various directions, setting $f=1$ and $\chi=1$, or $16$. For $\chi=1$, only the parallel steady state exists; to reach that state, the particle accelerates mainly in the direction of the force, with the initial transverse velocity gradually attenuating. For $\chi=16$, the parallel and the two anti-parallel states exist. For initial spontaneous motion collinear with the force, the particle approaches either the parallel or large-magnitude anti-parallel states. For initial spontaneous motion that is non-collinear with the force (regardless of the direction), the particle approaches the parallel state while remaining near the manifold  $v=\chi/16$ of unforced spontaneous-motion states. 

\begin{figure}[t!]
\begin{center}
\includegraphics[scale=0.4]{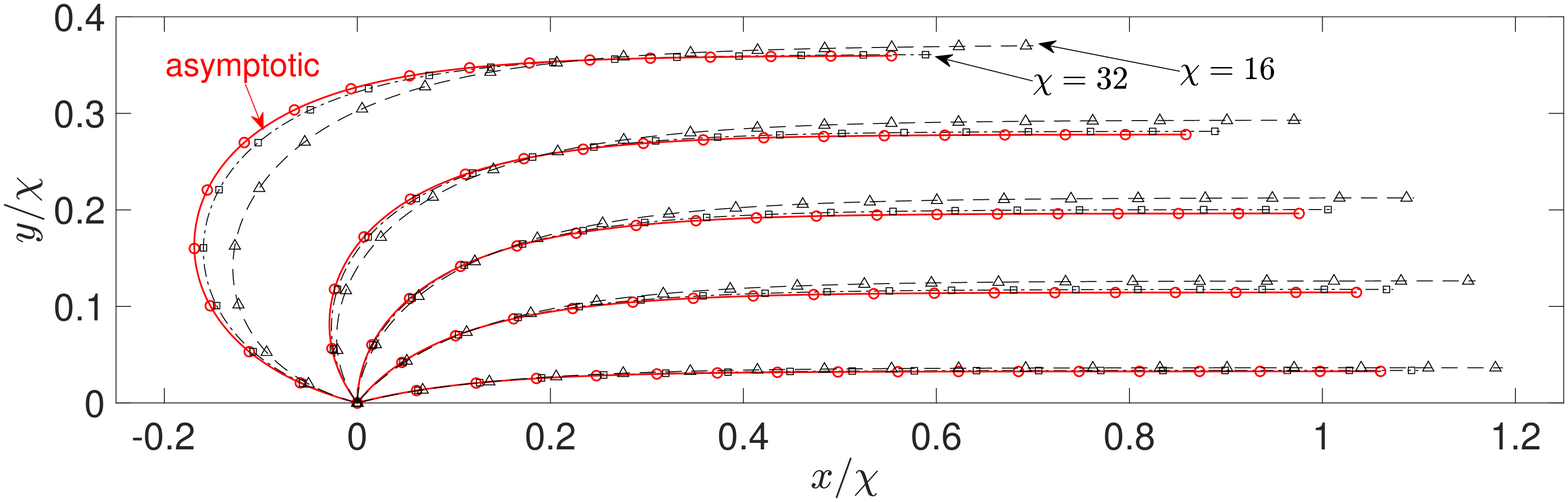}
\caption{Dynamical alignment of an isotropic active particle, initially at a steady-spontaneous state $\bar{\bv}=(\chi/16)\hat{\bp}$ with some direction $\hat{\bp}$ in the $x\,$--$\,y$ plane, subjected to a unit-step force in the $\be_x$ direction; the simulations are set up such that the particle is at the origin at time $t=0$ when the force starts to be applied. The black dashed and dash-dotted curves depict numerical simulations of particle position (scaled by $\chi$) for the indicated large values of $\chi$. The red solid curves depict the large-$\chi$ (equivalently, weak-force)  approximation developed in Sec.~\ref{ssec:alignment}. The symbols indicate unit time steps.}
\label{fig:slowalignment}
\end{center}
\end{figure}

To analyze the latter scenario analytically, it is convenient to consider the limit $\chi\gg1$ with $f$ fixed. 
Given our expectation that the particle velocity remains near the manifold of unforced spontaneous-motion states, we scale the velocity as $\bv(t)=\chi \bV(t)$. 
We further assume, subject to \textit{a posteriori} confirmation, that $\bV(t)$ varies on an order-unity time scale. In that case, we can use the `slowly varying' expansion \eqref{slow expansion} of the history operator (the time scale is long compared to $1/v^2=\text{ord}(1/\chi^2)$) to approximate the unsteady amplitude equation \eqref{amplitude repeat} as 
\begin{equation}\label{amplitude force approximated}
\chi^2(1-16V)\bV-\frac{4}{V}\left\{(\tI-\hat{\bV}\hat{\bV})\bcdot\frac{\mathrm{d}\bV}{\mathrm{d}t}+\mathcal{O}(\chi^{-4})\right\}+2f\unit=\bzero.
\end{equation}
We next expand  
\begin{equation}
\bV(t)\sim \bV_0(t) + \chi^{-2}\bV_1(t) \quad \text{as} \quad \chi\to+\infty.
\end{equation}
At the leading order $\chi^2$, \eqref{amplitude force approximated} gives 
\begin{equation}
(1-16|\bV_0|)\bV_0=\bzero,
\end{equation}
the general solution being
\begin{equation}\label{V0 sol}
\bV_0(t)=\frac{1}{16}\hat{\mathbf{P}}(t),
\end{equation}
wherein $\hat{\bP}(t)$ is a time-dependent unit vector. Namely, the particle velocity lies on the manifold of unforced spontaneous-motion states, with the direction of motion varying with time. At order unity, \eqref{amplitude force approximated} gives 
\begin{equation}\label{Phat eq}
\frac{\mathrm{d}{\hat{\bP}}}{\mathrm{d}t}=\frac{1}{2}f\unit\bcdot (\tI-\hat{\bP}\hat{\bP}),
\end{equation}
where we used the fact that $\mathrm{d}\hat{\bP}/\mathrm{d}t$ is perpendicular to $\hat{\bP}$ in order to eliminate the unknown correction $\bV_1(t)$. 

The motion is confined to the plane spanned by $\unit$ and the initial value of $\hat{\bP}$. Let $\vartheta(t)\in(0,\pi)$ be the angle between $\unit$ and $\hat{\bP}$ and introduce Cartesian coordinates $x\,$--$\,y$ such that $\bx= x\be_x+y\be_y$, with $\be_x=\unit$ and $\hat{\bP}=\cos\vartheta\be_x+\sin\vartheta\be_y$. Then   
\eqref{Phat eq} becomes 
\begin{equation}\label{th eq}
\frac{\mathrm{d}\vartheta}{\mathrm{d}t}=-\frac{f}{2}\sin\vartheta.
\end{equation}
By integrating from an initial angle $\vartheta_i$ to a final angle $\vartheta_f<\vartheta_i$ we obtain the alignment time 
\begin{equation}\label{dt slow}
\Delta t \sim \frac{2}{f}\ln\frac{\tan (\vartheta_i/2)}{\tan(\vartheta_f/2)}.
\end{equation}
Furthermore, integrating $\mathrm{d}\bx/\mathrm{d}t\sim \chi \bV_0$ using \eqref{V0 sol} and \eqref{th eq} yields the corresponding Cartesian displacements 
\refstepcounter{equation}\label{xy slow}
$$
 \Delta x\sim \frac{\chi}{8f}\ln\frac{\sin\vartheta_i}{\sin\vartheta_f}, \quad \Delta y \sim  \frac{\chi}{8f}(\vartheta_i-\vartheta_f).
 \eqno{(\theequation \mathrm{a},\mathrm{b})}
$$

The approximations \eqref{dt slow} and \eqref{xy slow} are illustrated and confirmed against numerical simulations in Fig.~\ref{fig:slowalignment}. While we have derived these approximations for $\chi\to+\infty$ with $f$ fixed, given the partial arbitrariness in the definition of $\chi$ they must hold as long as $\chi/\sqrt{f}\gg1$. Thus for weak force the alignment time is not order unity but long like $1/f$, whereas the displacements scale like $\chi/f$. As to be expected, the alignment time $\Delta t$ and longitudinal displacement $\Delta x$ diverge (logarithmically) as either $\vartheta_i\nearrow\pi$ or $\vartheta_f\searrow0$. In contrast, the transverse displacement $\Delta y$ limits to $\chi\vartheta_i/(8f)$ as $\vartheta_f\searrow0$, or $\chi\pi/(8f)$ when also $\vartheta_i\nearrow\pi$.

\section{Concluding remarks}
\label{sec:conclusions}
Starting from Michelin \textit{et al.}'s canonical model of an isotropic chemically active particle \cite{Michelin:13}, and including the possibility of a weak external force, we employed weakly nonlinear analysis to develop an unsteady, three-dimensional nonlinear amplitude equation \eqref{amplitude repeat} governing the slow dynamics of the particle valid in the limit where the P\'eclet number approaches its threshold value for instability and spontaneous motion in the force-free case. An unconventional feature of this unsteady amplitude equation is that it does not explicitly involve time derivatives; rather, unsteadiness enters via a nonlocal term---a time integral over the history of the particle motion representing the interaction of the particle with its own wake \eqref{H repeat}. This manifestation of unsteadiness is a consequence of the spatial non-uniformity of the weakly nonlinear expansion at large distances from the particle. Previously, this same non-uniformity has been linked to other unconventional features of the near-threshold dynamics, including the singular-pitchfork bifurcation in the unperturbed isotropic active-particle scenario and the fact that the amplitude equation arises from solvability at second, rather than third, order of the weakly nonlinear expansion \cite{Farutin:21}. 

In contrast to our unsteady theory, previous weakly nonlinear analyses of the canonical active-particle model and closely related models of `self-solubilizing' active drops have either been limited to steady states \cite{Rednikov:94,Rednikov:94b,Morozov:19,Morozov:19b,Morozov:19b,Saha:21} or employed an inconsistent `quasi-steady' assumption, as in the weakly nonlinear analysis of axisymmetric drop interaction in section 5 of \cite{Lippera:20} and the analysis of the stability of spontaneous-motion states in appendix B of \cite{Morozov:19b}. (Unsteady weakly nonlinear analyses of active particles and drops have been carried out before but starting from models of `reactive' active drops where there is no solute transport in the exterior bulk \cite{Rednikov:95}, or physically inconsistent active-particle models where solute transport is truncated at a finite distance from the particle \cite{Farutin:21b}; in those cases the weakly nonlinear expansion is spatially uniform and there is no history effect.) Our unsteady theory can also be related to the `moving singularity' active-drop model proposed  in \cite{Lippera:20b} where wake interaction is included via direct numerical simulation of an unsteady diffusion problem similar to our remote-region problem but with an additional dipolar source term which is meant to capture advection on the particle scale in an ad hoc manner. 

We employed our model to analyze and numerically demonstrate the stability characteristics of steady states for both force-free and forced particles. As to be expected from numerical simulations of the canonical active-particle model \cite{Michelin:13,Chen:21,Kailasham:22}, in the force-free case we found that above the threshold P\'eclet value the stationary state looses stability in favor of an isotropic manifold of spontaneous-motion states. In particular, the positive growth rates based on our three-dimensional stability analysis of the stationary state were found to agree asymptotically (approaching the threshold) with the axisymmetric linear stability analysis of the canonical model \cite{Michelin:13}. In the case of the spontaneous-motion states, there are no stability results in the literature to compare with. We have shown that these states are stable to longitudinal perturbations, as expected from axisymmetric numerical simulations \cite{Michelin:13,Kailasham:22}. We also showed that the spontaneous-motion states are only marginally stable to transverse perturbations, since transverse perturbations leave the particle velocity on the manifold of isotropic spontaneous-motion states. Nonlinearly, we observe that a spontaneously moving particle that is transversally perturbed indeed settles back into steady spontaneous motion but in a different direction. 

The forced case demonstrates well the need for analyses that are fully three-dimensional (rather than axisymmetric) as well as unsteady. It was previously shown for both active particles \cite{Saha:21} and active drops \cite{Rednikov:94,Rednikov:94b} that in the presence of a constant force, no matter how weak, steady translation is necessarily collinear with the force. Here, we have shown that the translational steady state parallel to the force that exists for all P\'eclet numbers (in the vicinity of the threshold) is stable. In contrast, the two anti-parallel steady states that exist above a critical P\'eclet value are both unstable, however the larger-magnitude anti-parallel state is stable under longitudinal perturbations. These findings agree with the heuristic `relative-stability' predictions in 
\cite{Rednikov:94,Rednikov:94b} for a closely related active-drop model, but contradict the guess made in \cite{Saha:21}, later referenced in \cite{Michelin:22} and \cite{Kailasham:22}, that the larger-magnitude anti-parallel state is stable. Motivated by the instability of the anti-parallel states, we have studied the alignment dynamics of a spontaneously moving particle following the sudden application of a force. In particular, we have derived an analytical description of the alignment in a limit where the effect of the force is weak and the particle velocity approximately traces the isotropic manifold of spontaneous-motion steady states from the force-free case. 

The unsteady theory developed herein builds upon the steady theory and adjoint formulation developed in part I of this series \cite{Schnitzer:22}. In particular, we used a special case of the general solvability condition derived therein to readily extract the amplitude equation for the particle velocity from the inhomogeneous problem at second order of the particle-scale weakly nonlinear expansion. By circumventing the need to solve that  inhomogeneous problem, which is generally not axisymmetric and depends on the details of the scenario being considered, we were able to develop a fully three-dimensional treatment and generalize the analysis in part I to the unsteady case without having to redo most of the quasi-static particle-scale analysis. 

As extensively demonstrated in part I in the steady case, the general solvability condition developed therein makes it straightforward to include in the modelling various perturbation effects, which generally have a leading-order effect sufficiently near the instability threshold. While in this part we have only included an external force as a perturbation, it would be natural to employ the general framework of part I in conjunction with the unsteady theory in this part towards exploring the dynamics of active particles in a range of perturbations scenarios. This would be immediate for perturbations that only modify the quasi-static particle-scale region. The perturbations could then be included as demonstrated in part I in the steady case, without any modification to the unsteady analysis of the remote region. This applies to most of the examples in part 1, including the external torque, surface-properties and surface-absorption perturbations. In other scenarios, the unsteady analysis of the remote region may require modification. For instance, we have seen in part I that weak bulk absorption can affect the remote region at leading order. In that scenario, the remote-region diffusion equation would include a reaction term proportional to a Damk\"ohler number characterizing the exponential decay of the concentration deviation and so the significance of history. 

The theory developed in this and the preceding part could also be readily applied to study $N$ interacting particles whose dimensionless separation is commensurate with the wake scale $1/\epsilon$. In the lab frame, the remote region would describe unsteady diffusion from $N$ moving sources of order-$\epsilon$ strength, the dominant mechanism of interaction being the order-$\epsilon^2$ concentration gradient induced on any given particle by the wakes of all other particles. We have seen that for an active particle an order-$\epsilon^2$ concentration gradient influences its leading order-$\epsilon$ velocity. In comparison, hydrodynamic interactions are negligible. Flat boundaries could also be included in the modelling using the method of images, simultaneously with other perturbations such as an external force. The only published weakly nonlinear analysis of interacting active particles (see section 5 of \cite{Lippera:20}) is limited to axisymmetric collisions and employs a quasi-static assumption that as already mentioned erroneously overlooks the history effect. This overlook and recent interest in the effects of interactions and boundaries on the motion of active drops \cite{Lippera:20b,Desai:21,Picella:22,Desai:22,Hokmabad:22} suggests this direction as an important application area for our theory. 

Lastly, recall that our theory is derived from the canonical active-particle model of Michelin \textit{et al.} \cite{Michelin:13}, which describes isotropic autocatalytic colloids but is often adopted as a conventional reference model for self-solubilizing active drops \cite{Michelin:22}. While autocatalytic colloids can be physically realized   \cite{Golestanian:05,Golestanian:07}, diffusion of the solute molecules they emit or absorb at their surface is typically too strong to allow for a hydro-chemical instability; namely, it is difficult in practice to reach the threshold P\'eclet number. Clearly, it would be desirable to generalize our theory to self-solubilizing active drops. Since drop deformation is usually negligible in these setups, the main differences are that the concentration surface gradient drives flow via a Marangoni effect rather than a diffusio-osmotic slip effect and that the flow in the drop interior needs to be considered. Since these differences can only affect the details of the particle-scale region, we expect that it  would be necessary to update the adjoint formulation developed in part I but not the analysis of the unsteady remote region carried out in this part. In light of the above, we anticipate amplitude equations having the same form as in the active-particle case but with different values of the coefficients. 

\textbf{Acknowledgments}. The authors are grateful to Mohit Dalwadi and an anonymous Referee for constructive comments and acknowledge the generous support of the Leverhulme Trust through Research Project Grant RPG-2021-161.

\appendix
\section{Local analysis of unsteady remote problem}\label{app:localanalysis}
In this appendix, we carry out a local analysis as $\tilde{r}\searrow0$ of the unsteady remote-region problem \eqref{remote eq}--\eqref{remote matching}. From this local analysis, we shall obtain the expansion \eqref{local behaviour} with the quantities $\tilde{h}(t)$ and $\tilde{\mathbf{H}}(t)$ undetermined.

Given the singular condition \eqref{remote matching}, we begin by writing 
 \begin{equation}\label{C0 def}
\tilde{c}(\tilde{\br},t)=\frac{1}{\tilde{r}}+C_0(\tilde{\br},t), \quad \text{with} \quad C_0(\tilde\br,t)=o(1/\tilde{r}) \quad \text{as} \quad \tilde{r}\searrow0.
 \end{equation}
Since $1/\tilde{r}$ satisfies Laplace's equation for $\tilde{r}>0$, the governing partial-differential equation \eqref{remote eq} gives 
\begin{equation}\label{C0 eq}
\pd{C_0}{t}-\bv\bcdot\tilde{\bnabla}\frac{1}{\tilde{r}}-\bv\bcdot\tilde{\bnabla} C_0 = \frac{1}{4}\tilde{\nabla}^2 C_0 \quad \text{for} \quad \tilde{r}>0. 
\end{equation}
Given \eqref{C0 def}, it is plausible to assume that
\begin{equation}\label{C0 assumptions}
\pd{C_0}{t},\tilde{\bnabla} C_0\ll\frac{1}{\tilde{r}^2} \quad \text{as} \quad \tilde{r}\searrow0,
\end{equation}  
in which case the dominant balance of \eqref{C0 eq} is 
\begin{equation}\label{C0 balance}
{\tilde\bnabla}^2C_0\sim -4\bv\bcdot\bnabla\frac{1}{\tilde{r}} \quad \text{as} \quad \tilde{r}\to0. 
\end{equation}
The balance \eqref{C0 balance} possesses the particular solution
\begin{equation}
-2\be_r\bcdot\bv.
\end{equation}
The only asymptotically consistent homogeneous solutions of \eqref{C0 balance}, namely solutions of Laplace's equation in $\tilde{r}>0$, are constant in space or harmonics increasing with $\tilde{r}$, the former dominating the latter as $\tilde{r}\searrow0$. We conclude that
\begin{equation}\label{C1 def}
C_0(\tilde{\br},t)=\tilde{h}(t)-2\be_r\bcdot\bv(t) + C_1(\tilde{\br},t),
\end{equation}
where $\tilde{h}(t)$ is constant in space and $C_1(\tilde{\br},t)=o(1)$ as $\tilde{r}\to0$. 

Using \eqref{C1 def} in \eqref{C0 eq}, we find 
\begin{equation}\label{C1 eq}
\frac{d\tilde{h}}{dt}-2\be_r\bcdot\frac{d\bv}{dt}+\pd{C_1}{t} +2\bv\bcdot \tilde{\bnabla}\left(\be_r\bcdot\bv\right)-\bv\bcdot\tilde{\bnabla} C_1=\frac{1}{4}\tilde{\nabla}^2 C_1 \quad \text{for} \quad \tilde{r}>0. 
\end{equation}
Making the plausible assumption that $\tilde{\bnabla} C_1\ll 1/\tilde{r}$ as $\tilde{r}\searrow0$, the left-hand side of \eqref{C1 eq} is dominated by the order-$1/\tilde{r}$ fourth term. Simplifying that term, we find the dominant balance 
\begin{equation}\label{C1 balance}
\tilde{\nabla}^2C_1
\sim \frac{8}{\tilde{r}}\bv\bv\boldsymbol{:}(\tI-\be_r\be_r) \quad \text{as} \quad \tilde{r}\searrow0. 
\end{equation}
The balance \eqref{C1 balance} possesses  the particular solution 
\begin{equation}
2\tilde{r}\bv\bv\boldsymbol{:}(\tI+\be_r\be_r).
\end{equation}
The only asymptotically consistent homogeneous solutions to \eqref{C1 balance} are growing harmonics, the dominant of which can be written as $\tilde{\mathbf{H}}\bcdot \bX$, with $\tilde{\mathbf{H}}$ a spatially constant vector. We therefore obtain the local expansion
\begin{equation}
\tilde{c}=\frac{1}{\tilde{r}}+\tilde{h}(t)-2\bv\bcdot\be_r+\tilde{r}\left\{2\bv\bv\boldsymbol{:}(\tI+\be_r\be_r)+\tilde{\mathbf{H}}(t)\bcdot\be_r\right\} + o(\tilde{r}) \quad \text{as} \quad \tilde{r}\searrow0,
\end{equation}
which is the same as \eqref{local behaviour} in the main text. 

\section{Steady motion for $t<0$}
\label{app:tailearly}
\subsection{Analytical integration over the history operator's tail}
\label{app_s:tail}
Consider the scenario where $\bv(t)=\bar{\bv}$ for $t<0$, where $\bar{\bv}$ is a steady state  (magnitude $\bar{v}$) for some constant force $\mathbf{f}_-$ such that (cf.~\eqref{amplitude repeat})
\begin{equation}\label{steady before}
(\chi-16\bar{v})\bar{\bv}+2\mathbf{f}_-=\bzero.
\end{equation}
For $t>0$, we split the history operator \eqref{H def} as
\begin{equation}
\mathbf{H}[\bv](t)=\mathbf{H}\ub{-}[\bv(\tau)]_{\tau=0}^t(t;\bar{\bv})
+\mathbf{H}\ub{+}[\bv(\tau)]_{\tau=0}^t(t;\bar{\bv}),
\end{equation}
where the `tail' contribution from past times $\tau<0$ is 
\begin{equation}\label{history tail}
\mathbf{H}\ub{-}[\bv(\tau)]_{\tau=0}^t(t;\bar{\bv})=\frac{2}{\sqrt{\pi}}\int_{-\infty}^0\frac{\mathrm{d}\tau}{(t-\tau)^{5/2}}\left\{(t-\tau)\bv(t)e^{{-(t-\tau)v^2(t)}}-\bx'(\tau,t)e^{-\tfrac{|\bx'(\tau,t)|^2}{t-\tau}}\right\},
\end{equation}
and the contribution from positive past times $0<\tau<t$ is 
\begin{equation}\label{history recent}
\mathbf{H}\ub{+}[\bv(\tau)]_{\tau=0}^t(t;\bar{\bv})=\frac{2}{\sqrt{\pi}}\int_{0}^t\frac{\mathrm{d}\tau}{(t-\tau)^{5/2}}\left\{(t-\tau)\bv(t)e^{{-(t-\tau)v^2(t)}}-\bx'(\tau,t)e^{-\tfrac{|\bx'(\tau,t)|^2}{t-\tau}}\right\}.
\end{equation}
We wish to analytically evaluate the tail contribution. 

Recalling the definition \eqref{xp def}, we have 
\begin{equation}\label{xp tail}
\bx'(\tau,t)=\bx(t)-\bx(0)-\tau\bar{\bv},
\end{equation}
and so 
\begin{equation}\label{xpxp tail}
\frac{\bx'(\tau,t)\bcdot \bx'(\tau,t)}{t-\tau}=\bar{v}^2(t-\tau)+2\bar{\bv}\bcdot\delta\bx(t)+\frac{\delta x^2(t)}{t-\tau},
\end{equation}
where we define 
\begin{equation}\label{deltax}
\delta\bx(t)=\bx(t)-\bx(0)-\bar{\bv}t
\end{equation} 
and $\delta x(t)=|\delta\bx(t)|$. Upon making the change of variables $p=t-\tau$, \eqref{history tail} gives, using \eqref{xp tail} and \eqref{xpxp tail}, 
\begin{multline}
\mathbf{H}\ub{-}[\bv(\tau)]_{\tau=0}^t(t;\bar{\bv})=\frac{2\bv(t)}{\sqrt{\pi}}\int_t^{\infty}\frac{\mathrm{d}p}{p^{3/2}}e^{-v^2(t)p}\\
-\frac{2}{\sqrt{\pi}}e^{-2\bar{\bv}\bcdot \delta\bx(t)}\left\{\delta\bx(t)\int_t^{\infty}\frac{\mathrm{d}p}{p^{5/2}}e^{-\bar{v}^2p-\delta x^2(t)/p}+\bar{\bv}\int_t^{\infty}\frac{\mathrm{d}p}{p^{3/2}}e^{-\bar{v}^2p-\delta x^2(t)/p}\right\}.
\end{multline}
Performing the quadratures, we find
\begin{multline}\label{tail analytical}
\mathbf{H}\ub{-}[\bv(\tau)]_{\tau=0}^t(t;\bar{\bv})=4\bv(t)\left\{\frac{e^{-tv^2(t)}}{(\pi t)^{1/2}}-v(t)\,\mathrm{erfc}[t^{1/2}v(t)]\right\}\\
-\frac{2}{\sqrt{\pi}}e^{-2\bar{\bv}\bcdot\delta\bx(t)}\left\{-\frac{\delta\bx(t)}{t^{1/2}\delta x^2(t)}e^{-\bar{v}^2t-\delta x^2(t)/t} \right.\\ \left.
+\frac{\sqrt{\pi}}{4\delta x^3(t)}e^{-2\bar{v}\delta x(t)}\left[\delta\bx(t)+2\bar{v}\delta\bx(t)\delta x(t)+2\bar{\bv}\delta x^2(t)\right]\,\mathrm{erfc}\frac{t\bar{v}-\delta x(t)}{t^{1/2}} \right.\\\left.
-\frac{\sqrt{\pi}}{4\delta x^3(t)}e^{2\bar{v}\delta x(t)}\left[\delta\bx(t)-2\bar{v}\delta x(t)\delta\bx(t)+2\bar{\bv}\delta x^2(t)\right]\,\mathrm{erfc}\frac{t\bar{v}+\delta x(t)}{t^{1/2}}\right\}.
\end{multline}

\subsection{Early-time asymptotic expansion}
\label{app_s:early}
Consider now the early-time limit $t\searrow0$ of the particle velocity $\bv(t)$ in the same scenario as described above \eqref{steady before}, with the force $\mathbf{f}(t)$ assumed to be smooth for $t>0$, with a finite limit $\mathbf{f}(t)\to \mathbf{f}_+$ as $t\searrow0$ that may differ from the constant force value $\mathbf{f}_-$ corresponding to the steady state $\bv(t)=\bar{\bv}$ for $t<0$ (cf.~\eqref{steady before}). 

We claim that the particle velocity possesses an expansion of the form 
\begin{equation}\label{early v expansion}
\bv(t)=\bar{\bv}+t^{1/2}\bv_1+\mathcal{O}(t) \quad \text{as} \quad t\searrow0,
\end{equation}
where the fractional powers are suggested by the form of the history operator \eqref{H repeat} and we have assumed that the velocity is continuous at $t=0$. It can be shown that without the latter assumption the history operator is asymptotically large as $t\searrow0$, with no other comparable term in the amplitude equation \eqref{amplitude repeat} to balance it. 

Consider now the history operator $\mathbf{H}[\bv](t)$ as $t\searrow0$. Carefully expanding the closed-form expression for the tail contribution \eqref{tail analytical} gives, 
\begin{equation}\label{early T cont}
\mathbf{H}\ub{-}[\bv(\tau)]_{\tau=0}^t(t;\bar{\bv})=\frac{28}{9\sqrt{\pi}}\bv_1+\mathcal{O}(t^{1/2})\quad \text{as} \quad t\searrow0,
\end{equation}
where we have used \eqref{deltax} and \eqref{early v expansion} to obtain $\delta\bx = 2t^{3/2}\bv_1/3+\mathcal{O}(t^2)$ in the same limit. Consider next the contribution \eqref{history recent} from positive past times. Upon making the change of variables $p=t-\tau$, we have
\begin{equation}\label{early Hp int}
\mathbf{H}\ub{+}[\bv(\tau)]_{\tau=0}^t(t;\bar{\bv})=\frac{2}{\sqrt{\pi}}\int_{0}^t\frac{\mathrm{d}p}{p^{5/2}}\left\{p\bv(t)e^{{-pv^2(t)}}-\bx'(t-p,t)e^{-\tfrac{|\bx'(t-p,t)|^2}{p}}\right\}.
\end{equation}
Noting that (cf.~\eqref{xp def})
\begin{equation}
\bx'(t-p,t)=\bar{\bv} p +\frac{2}{3}\bv_1\left[t^{3/2}-(t-p)^{3/2}\right]+\mathcal{O}(t^2) \quad \text{as} \quad t\searrow0,
\end{equation}
a straightforward expansion of \eqref{early Hp int} gives
\begin{equation}\label{early p cont}
\mathbf{H}\ub{+}[\bv(\tau)]_{\tau=0}^t(t;\bar{\bv})=\frac{4(3\pi-7)}{9\sqrt{\pi}}\bv_1+\mathcal{O}(t^{1/2})\quad \text{as} \quad t\searrow0.
\end{equation}
Combining the contributions \eqref{early T cont} and \eqref{early p cont} then gives
\begin{equation}\label{H early}
\mathbf{H}[\bv]= \frac{4\sqrt{\pi}}{3}\bv_1+\mathcal{O}(t^{1/2}) \quad \text{as} \quad t\searrow0.
\end{equation}

Using \eqref{early v expansion} and \eqref{H early} to expand the amplitude equation \eqref{amplitude repeat}, we find
\begin{equation}
\bv_1=\frac{3}{16\sqrt{\pi}}\left[\left(\chi-16\bar{v}\right)\bar{\bv}+2\mathbf{f}_+\right].
\end{equation}
Clearly, $\bv_1$ vanishes if and only if 
the force is continuous at $t=0$, that is if $\mathbf{f}_{+}=\mathbf{f}_-$.  

\section{Numerical regularization}
\label{app:reg}
In our numerical scheme, the contribution \eqref{history recent} of positive past times is evaluated by the trapezoidal rule. The integrable $1/(t-\tau)^{1/2}$ singularity of the integrand (cf.~\eqref{integrand expansion} below) induces a $\mathcal{O}(\Delta t^{1/2})$ discretization error, where $\Delta t$ is the time step. To improve on this, we note that for $t>0$ we have 
\begin{equation}
\bx'(\tau,t)=(t-\tau)\bv(t)-\frac{1}{2}(t-\tau)^2\ba(t) + \mathcal{O}\left((t-\tau)^3)\right) \quad \text{as} \quad \tau\nearrow t, 
\end{equation}
where $\ba(t)=\mathrm{d}\bv(t)/\mathrm{d}t$ is the acceleration. It follows that in the same limit
\begin{equation}
-\bx'(\tau,t)e^{-\frac{|\bx'(\tau,t)|^2}{t-\tau}}=e^{-(t-\tau)v^2(t)}\left\{-(t-\tau)\bv(t)+\frac{1}{2}(t-\tau)^2\ba(t)\right\} + \mathcal{O}\left((t-\tau)^3)\right),
\end{equation}
where for later convenience we do not expand the exponent on the right-hand side. Thus, the integrand in \eqref{history recent} possesses the expansion
\begin{equation}\label{integrand expansion}
\frac{\mathbf{a}(t)}{(t-\tau)^{1/2}}e^{-(t-\tau)v^2(t)}+\mathcal{O}\left((t-\tau)^{1/2}\right) \quad \text{as} \quad \tau\nearrow t
\end{equation}
and so \eqref{history recent} can be regularized by subtracting and adding the integral 
\begin{equation}
\frac{\ba(t)}{\sqrt{\pi}}\int_0^t\frac{\mathrm{d}\tau}{(t-\tau)^{1/2}}e^{-(t-\tau)v^2(t)}=\ba(t)\frac{\mathrm{erf}(t^{1/2}v^{1/2}(t))}{v^{1/2}(t)}.
\end{equation}
This reduces the order of the discretization error to $\Delta t^{3/2}$ with the small penalty of having to estimate the acceleration $\ba(t)$ as part of the numerical scheme. 

\section{Linear impulse response}
\label{app:impulse}

The linear stability analysis in Sec.~\ref{ssec:stability} assumes an exponential time dependence $\delta\bv = \mathbf{A} e^{\sigma t}$, and results in equations \eqref{sigma stat} and \eqref{sigma prime eq} for the growth rate $\sigma$ which can be written in the form $F(\sigma) = 0$, where
\begin{gather}\label{sigma F eq}
	F(\sigma) = -\chi + \frac{32 \sigma^{1/2}}{3} \quad \text{and} \quad {-\chi} + \frac{32}{3}\left[\frac{(\bar{v}^2+ \sigma)^{3/2} - \bar{v}^3}{\sigma} + 3 \delta_\parallel \bar{v}^2\frac{(\bar{v}^2 + \sigma)^{1/2} - \bar{v}}{\sigma}\right],
\end{gather}
respectively, for the static base state with $\bar{\bv} = \bzero$ and for the steady rectilinear-motion base state with $\bar{\bv} \neq \bzero$ (and $\delta_\parallel = 1$ for $\mathbf{A} \parallel \bar{\bv}$ or $\delta_\parallel = 0$ for $\mathbf{A} \perp \bar{\bv}$). These equations are only valid for $\mathop{\mathrm{Re}} \sigma \geq -\bar{v}^2$, as the linearized history integral \eqref{dH basic} diverges otherwise. Here, we seek to calculate the decay rates of the perturbations when there is no such solution $\sigma$.

We do this by considering the linear response to a force impulse of amplitude $\mathbf{B}$, i.e.~the solution $\delta\bv(t)$ to
\begin{gather}
	\left\{\chi-16\bar{v}\left(\tI+\hat{\bar{\bv}}\hat{\bar{\bv}}\right)\right\}\bcdot\delta\bv
-4\,\delta\mathbf{H}[\delta\bv;\bar{\bv}] = 2 \mathbf{B} \delta(t)
\end{gather}
that satisfies $\delta\bv(t) = 0$ for $t < 0$, where $\delta(t)$ is the Dirac delta function. Using a Laplace transform,
\newcommand{\LT}[1]{\mathcal{L}\{#1\}}
\begin{gather}
	\LT{\delta \mathbf{v}}(s) = \int_{0^-}^\infty \mathrm{d}t\,\delta\mathbf{v}(t) e^{st},
\end{gather}
and evaluating \eqref{H linear zero} or \eqref{dH def steady motion}, we obtain the result
\begin{gather}
	F(s)\LT{\delta\bv} = 2 \mathbf{B},
\end{gather}
where the function $F(s)$ is given by \eqref{sigma F eq}. The solution is then given by the Laplace inversion integral 
\begin{gather}
	\delta\bv(t) = \frac{\mathbf{B}}{\pi i} \int_{c-i\infty}^{c+i\infty} \mathrm{d}s\,\frac{e^{s t}}{F(s)},
\end{gather}
where the contour passes to the right of any singularities of the integrand. We determine the late-time behaviour of this integral using the method of steepest descent. The calculation of $F(s)$ requires $\mathop{\mathrm{Re}} s \geq -\bar{v}^2$ for the history integral to converge, so we extend the results \eqref{sigma F eq} analytically, placing the branch cut for the square roots on the real axis where $s < -\bar{v}^2$. A suitable steepest-descent contour then runs just below the branch cut from $-\infty$ to the branch point $-\bar{v}^2$, anticlockwise around the branch point, and then just above the branch cut back to $-\infty$. 

We first consider the static base state, $\bar{\bv} = \bzero$. For $\chi < 0$, the equation $F(\sigma) = 0$ has no solutions, so we can deform onto the steepest-descent contour without crossing any singularities of the integrand. We then use the parametrization $s = -x$ to obtain
\begin{gather}\label{sd static}
	\delta\bv =\frac{\mathbf{B}}{\pi i} \int_0^\infty \mathrm{d}x \left[\frac{-1}{\chi + i\tfrac{32}{3} x^{1/2}} + \frac{1}{\chi - i\tfrac{32}{3} x^{1/2}}\right]e^{-x t} 
	=
	\frac{64 \mathbf{B}}{3\pi} \int_0^\infty \mathrm{d}x\,\frac{x^{1/2}\,e^{-x t}}{\chi^2 + \tfrac{1024}{9} x},
\end{gather}
from which Watson's lemma yields the leading-order behaviour
\begin{gather}
	\delta\bv \sim \frac{32 \mathbf{B}}{3\sqrt{\pi} \chi^2} t^{-3/2} \quad \text{as} \quad t \to \infty.
\end{gather}
For $\chi = 0$, the same contour yields an exact result
\begin{gather}
	\delta\bv = \frac{3 \mathbf{B}}{16\pi}\int_0^\infty \mathrm{d}x\, \frac{e^{-xt}}{x^{1/2}} = \frac{3 \mathbf{B}}{16\sqrt{\pi}} t^{-1/2}.
\end{gather}
For $\chi > 0$, the contribution from the contour is again given by \eqref{sd static}, but the contour deformation crosses a pole, located at the solution $s = \sigma$ of $F(\sigma) = 0$, and picks up a residue contribution (that is exponentially dominant over the contour contribution), so that
\begin{gather}
	\delta\bv \sim \frac{2\mathbf{B} e^{\sigma t}}{F'(\sigma)} = \frac{3 \mathbf{B} \sigma^{1/2} e^{\sigma t}}{8},
\end{gather}
which we recognize as the exponential behaviour predicted in Sec.~\ref{ssec:stability}, but with amplitude determined by the impulse amplitude. 

For steady rectilinear motion $\bar{\bv} \neq \bzero$, the calculation proceeds in the same way. Using the parametrization $s = -\bar{v}^2(1+x)$, we express the steepest-descent integral as
\begin{gather}
\frac{64\mathbf{B} \bar{v} e^{-\bar{v}^2 t}}{3\pi} \int_0^\infty \mathrm{d}x\,\frac{(1+x)(x - 3 \delta_\parallel)x^{1/2}}{[(1+x)\chi/\bar{v} - \tfrac{32}{3}(1+3 \delta_\parallel)]^2 + \tfrac{1024}{9} x(3 \delta_\parallel-x)^2} e^{-x \bar{v}^2 t}.
\end{gather}
Thus, when $F(\sigma) = 0$ has no solutions with $\mathop{\mathrm{Re}} \sigma \geq -\bar{v}^2$, Watson's lemma yields the leading-order results, as $t \to \infty$,
\begin{gather}
	\delta\bv \sim -\frac{32 \mathbf{B}}{\sqrt{\pi} (\chi - \tfrac{128}{3}\bar{v})^2} e^{-\bar{v}^2 t} t^{-3/2}
	\quad \text{or} \quad 
	\delta\bv \sim \frac{16 \mathbf{B} \bar{v}^{-2}}{\sqrt{\pi}(\chi - \tfrac{32}{3}\bar{v})^2} e^{-\bar{v}^2 t}t^{-5/2},
\end{gather}
depending on whether the perturbation is longitudinal or transverse, respectively. If, on the other hand, there are such solutions $\sigma$, then the associated residue contributions are dominant, yielding exponential solutions (except that for $\chi = 32\bar{v}$ the solution $\sigma = 0$ is a double root, leading to a perturbation of the form $\delta\bv \sim at + b$). 

We conclude that, for all three cases, when the growth-rate equation $F(\sigma) = 0$ has a solution with $\mathop{\mathrm{Re}} \sigma > -\bar{v}^2$, then there are perturbations with exponential time dependence $e^{\sigma t}$. If there are no such solutions, then the perturbations decay like $e^{-\bar{v}^2 t}$ times a power of $t$, and hence the base state is stable, as was asserted in Sec.~\ref{ssec:stability}.

\bibliography{refs}

\begin{thebibliography}{28}%
\makeatletter
\providecommand \@ifxundefined [1]{%
 \@ifx{#1\undefined}
}%
\providecommand \@ifnum [1]{%
 \ifnum #1\expandafter \@firstoftwo
 \else \expandafter \@secondoftwo
 \fi
}%
\providecommand \@ifx [1]{%
 \ifx #1\expandafter \@firstoftwo
 \else \expandafter \@secondoftwo
 \fi
}%
\providecommand \natexlab [1]{#1}%
\providecommand \enquote  [1]{``#1''}%
\providecommand \bibnamefont  [1]{#1}%
\providecommand \bibfnamefont [1]{#1}%
\providecommand \citenamefont [1]{#1}%
\providecommand \href@noop [0]{\@secondoftwo}%
\providecommand \href [0]{\begingroup \@sanitize@url \@href}%
\providecommand \@href[1]{\@@startlink{#1}\@@href}%
\providecommand \@@href[1]{\endgroup#1\@@endlink}%
\providecommand \@sanitize@url [0]{\catcode `\\12\catcode `\$12\catcode
  `\&12\catcode `\#12\catcode `\^12\catcode `\_12\catcode `\%12\relax}%
\providecommand \@@startlink[1]{}%
\providecommand \@@endlink[0]{}%
\providecommand \url  [0]{\begingroup\@sanitize@url \@url }%
\providecommand \@url [1]{\endgroup\@href {#1}{\urlprefix }}%
\providecommand \urlprefix  [0]{URL }%
\providecommand \Eprint [0]{\href }%
\providecommand \doibase [0]{http://dx.doi.org/}%
\providecommand \selectlanguage [0]{\@gobble}%
\providecommand \bibinfo  [0]{\@secondoftwo}%
\providecommand \bibfield  [0]{\@secondoftwo}%
\providecommand \translation [1]{[#1]}%
\providecommand \BibitemOpen [0]{}%
\providecommand \bibitemStop [0]{}%
\providecommand \bibitemNoStop [0]{.\EOS\space}%
\providecommand \EOS [0]{\spacefactor3000\relax}%
\providecommand \BibitemShut  [1]{\csname bibitem#1\endcsname}%
\let\auto@bib@innerbib\@empty
\bibitem [{\citenamefont {Michelin}\ \emph {et~al.}(2013)\citenamefont
  {Michelin}, \citenamefont {Lauga},\ and\ \citenamefont
  {Bartolo}}]{Michelin:13}%
  \BibitemOpen
  \bibfield  {author} {\bibinfo {author} {\bibfnamefont {S.}~\bibnamefont
  {Michelin}}, \bibinfo {author} {\bibfnamefont {E.}~\bibnamefont {Lauga}}, \
  and\ \bibinfo {author} {\bibfnamefont {D.}~\bibnamefont {Bartolo}},\
  }\bibfield  {title} {\enquote {\bibinfo {title} {Spontaneous autophoretic
  motion of isotropic particles},}\ }\href@noop {} {\bibfield  {journal}
  {\bibinfo  {journal} {Phys. Fluids}\ }\textbf {\bibinfo {volume} {25}},\
  \bibinfo {pages} {061701} (\bibinfo {year} {2013})}\BibitemShut {NoStop}%
\bibitem [{\citenamefont {Golestanian}\ \emph {et~al.}(2005)\citenamefont
  {Golestanian}, \citenamefont {Liverpool},\ and\ \citenamefont
  {Ajdari}}]{Golestanian:05}%
  \BibitemOpen
  \bibfield  {author} {\bibinfo {author} {\bibfnamefont {R.}~\bibnamefont
  {Golestanian}}, \bibinfo {author} {\bibfnamefont {T.~B.}\ \bibnamefont
  {Liverpool}}, \ and\ \bibinfo {author} {\bibfnamefont {A.}~\bibnamefont
  {Ajdari}},\ }\bibfield  {title} {\enquote {\bibinfo {title} {Propulsion of a
  molecular machine by asymmetric distribution of reaction products},}\
  }\href@noop {} {\bibfield  {journal} {\bibinfo  {journal} {Phys. Rev. Lett.}\
  }\textbf {\bibinfo {volume} {94}},\ \bibinfo {pages} {220801} (\bibinfo
  {year} {2005})}\BibitemShut {NoStop}%
\bibitem [{\citenamefont {Golestanian}\ \emph {et~al.}(2007)\citenamefont
  {Golestanian}, \citenamefont {Liverpool},\ and\ \citenamefont
  {Ajdari}}]{Golestanian:07}%
  \BibitemOpen
  \bibfield  {author} {\bibinfo {author} {\bibfnamefont {R.}~\bibnamefont
  {Golestanian}}, \bibinfo {author} {\bibfnamefont {T.~B.}\ \bibnamefont
  {Liverpool}}, \ and\ \bibinfo {author} {\bibfnamefont {A.}~\bibnamefont
  {Ajdari}},\ }\bibfield  {title} {\enquote {\bibinfo {title} {Designing
  phoretic micro-and nano-swimmers},}\ }\href@noop {} {\bibfield  {journal}
  {\bibinfo  {journal} {New J. Phys.}\ }\textbf {\bibinfo {volume} {9}},\
  \bibinfo {pages} {126} (\bibinfo {year} {2007})}\BibitemShut {NoStop}%
\bibitem [{\citenamefont {Michelin}(2022)}]{Michelin:22}%
  \BibitemOpen
  \bibfield  {author} {\bibinfo {author} {\bibfnamefont {S.}~\bibnamefont
  {Michelin}},\ }\bibfield  {title} {\enquote {\bibinfo {title}
  {Self-propulsion of chemically active droplets},}\ }\href@noop {} {\bibfield
  {journal} {\bibinfo  {journal} {Ann. Rev. Fluid Mech.}\ }\textbf {\bibinfo
  {volume} {55}} (\bibinfo {year} {2022})}\BibitemShut {NoStop}%
\bibitem [{\citenamefont {Schnitzer}(in press)}]{Schnitzer:22}%
  \BibitemOpen
  \bibfield  {author} {\bibinfo {author} {\bibfnamefont {O.}~\bibnamefont
  {Schnitzer}},\ }\bibfield  {title} {\enquote {\bibinfo {title} {Weakly
  nonlinear dynamics of a chemically active particle near the threshold for
  spontaneous motion. {I}. adjoint method},}\ }\href@noop {} {\bibfield
  {journal} {\bibinfo  {journal} {Phys. Rev. Fluids}\ } (\bibinfo {year} {in
  press})}\BibitemShut {NoStop}%
\bibitem [{\citenamefont {Rednikov}\ \emph
  {et~al.}(1994{\natexlab{a}})\citenamefont {Rednikov}, \citenamefont
  {Ryazantsev},\ and\ \citenamefont {Velarde}}]{Rednikov:94}%
  \BibitemOpen
  \bibfield  {author} {\bibinfo {author} {\bibfnamefont {A.~Y.}\ \bibnamefont
  {Rednikov}}, \bibinfo {author} {\bibfnamefont {Y.~S.}\ \bibnamefont
  {Ryazantsev}}, \ and\ \bibinfo {author} {\bibfnamefont {M.~G.}\ \bibnamefont
  {Velarde}},\ }\bibfield  {title} {\enquote {\bibinfo {title} {Drop motion
  with surfactant transfer in a homogeneous surrounding},}\ }\href@noop {}
  {\bibfield  {journal} {\bibinfo  {journal} {Phys. Fluids}\ }\textbf {\bibinfo
  {volume} {6}},\ \bibinfo {pages} {451--468} (\bibinfo {year}
  {1994}{\natexlab{a}})}\BibitemShut {NoStop}%
\bibitem [{\citenamefont {Rednikov}\ \emph
  {et~al.}(1994{\natexlab{b}})\citenamefont {Rednikov}, \citenamefont
  {Ryazantsev},\ and\ \citenamefont {Vel{\'a}rde}}]{Rednikov:94b}%
  \BibitemOpen
  \bibfield  {author} {\bibinfo {author} {\bibfnamefont {A.~Y.}\ \bibnamefont
  {Rednikov}}, \bibinfo {author} {\bibfnamefont {Y.~S.}\ \bibnamefont
  {Ryazantsev}}, \ and\ \bibinfo {author} {\bibfnamefont {G.}~\bibnamefont
  {Vel{\'a}rde}, \bibfnamefont {M}},\ }\bibfield  {title} {\enquote {\bibinfo
  {title} {Drop motion with surfactant transfer in an inhomogeneous medium},}\
  }\href@noop {} {\bibfield  {journal} {\bibinfo  {journal} {Int. J. Heat Mass
  Transf.}\ }\textbf {\bibinfo {volume} {37}},\ \bibinfo {pages} {361--374}
  (\bibinfo {year} {1994}{\natexlab{b}})}\BibitemShut {NoStop}%
\bibitem [{\citenamefont {Morozov}\ and\ \citenamefont
  {Michelin}(2019{\natexlab{a}})}]{Morozov:19}%
  \BibitemOpen
  \bibfield  {author} {\bibinfo {author} {\bibfnamefont {M.}~\bibnamefont
  {Morozov}}\ and\ \bibinfo {author} {\bibfnamefont {S.}~\bibnamefont
  {Michelin}},\ }\bibfield  {title} {\enquote {\bibinfo {title} {Nonlinear
  dynamics of a chemically-active drop: From steady to chaotic
  self-propulsion},}\ }\href@noop {} {\bibfield  {journal} {\bibinfo  {journal}
  {J. Chem. Phys.}\ }\textbf {\bibinfo {volume} {150}},\ \bibinfo {pages}
  {044110} (\bibinfo {year} {2019}{\natexlab{a}})}\BibitemShut {NoStop}%
\bibitem [{\citenamefont {Morozov}\ and\ \citenamefont
  {Michelin}(2019{\natexlab{b}})}]{Morozov:19b}%
  \BibitemOpen
  \bibfield  {author} {\bibinfo {author} {\bibfnamefont {M.}~\bibnamefont
  {Morozov}}\ and\ \bibinfo {author} {\bibfnamefont {S.}~\bibnamefont
  {Michelin}},\ }\bibfield  {title} {\enquote {\bibinfo {title}
  {Self-propulsion near the onset of marangoni instability of deformable active
  droplets},}\ }\href@noop {} {\bibfield  {journal} {\bibinfo  {journal} {J.
  Fluid Mech.}\ }\textbf {\bibinfo {volume} {860}},\ \bibinfo {pages}
  {711--738} (\bibinfo {year} {2019}{\natexlab{b}})}\BibitemShut {NoStop}%
\bibitem [{\citenamefont {Saha}\ \emph {et~al.}(2021)\citenamefont {Saha},
  \citenamefont {Yariv},\ and\ \citenamefont {Schnitzer}}]{Saha:21}%
  \BibitemOpen
  \bibfield  {author} {\bibinfo {author} {\bibfnamefont {S.}~\bibnamefont
  {Saha}}, \bibinfo {author} {\bibfnamefont {E.}~\bibnamefont {Yariv}}, \ and\
  \bibinfo {author} {\bibfnamefont {O.}~\bibnamefont {Schnitzer}},\ }\bibfield
  {title} {\enquote {\bibinfo {title} {Isotropically active colloids under
  uniform force fields: from forced to spontaneous motion},}\ }\href@noop {}
  {\bibfield  {journal} {\bibinfo  {journal} {J. Fluid Mech.}\ }\textbf
  {\bibinfo {volume} {916}} (\bibinfo {year} {2021})}\BibitemShut {NoStop}%
\bibitem [{\citenamefont {Farutin}\ and\ \citenamefont
  {Misbah}(2021)}]{Farutin:21}%
  \BibitemOpen
  \bibfield  {author} {\bibinfo {author} {\bibfnamefont {A.}~\bibnamefont
  {Farutin}}\ and\ \bibinfo {author} {\bibfnamefont {C.}~\bibnamefont
  {Misbah}},\ }\bibfield  {title} {\enquote {\bibinfo {title} {Singular
  bifurcations: a regularization theory},}\ }\href@noop {} {\bibfield
  {journal} {\bibinfo  {journal} {arXiv preprint arXiv:2112.12094}\ } (\bibinfo
  {year} {2021})}\BibitemShut {NoStop}%
\bibitem [{\citenamefont {Yariv}\ and\ \citenamefont
  {Kaynan}(2017)}]{Yariv:17:Uri}%
  \BibitemOpen
  \bibfield  {author} {\bibinfo {author} {\bibfnamefont {E.}~\bibnamefont
  {Yariv}}\ and\ \bibinfo {author} {\bibfnamefont {U.}~\bibnamefont {Kaynan}},\
  }\bibfield  {title} {\enquote {\bibinfo {title} {Phoretic drag reduction of
  chemically active homogeneous spheres under force fields and shear flows},}\
  }\href@noop {} {\bibfield  {journal} {\bibinfo  {journal} {Phys. Rev.
  Fluids}\ }\textbf {\bibinfo {volume} {2}},\ \bibinfo {pages} {012201}
  (\bibinfo {year} {2017})}\BibitemShut {NoStop}%
\bibitem [{\citenamefont {Kailasham}\ and\ \citenamefont
  {Khair}(2022)}]{Kailasham:22}%
  \BibitemOpen
  \bibfield  {author} {\bibinfo {author} {\bibfnamefont {R.}~\bibnamefont
  {Kailasham}}\ and\ \bibinfo {author} {\bibfnamefont {A.~S.}\ \bibnamefont
  {Khair}},\ }\bibfield  {title} {\enquote {\bibinfo {title} {Dynamics of
  forced and unforced autophoretic particles},}\ }\href@noop {} {\bibfield
  {journal} {\bibinfo  {journal} {J. Fluid Mech.}\ }\textbf {\bibinfo {volume}
  {948}},\ \bibinfo {pages} {A41} (\bibinfo {year} {2022})}\BibitemShut
  {NoStop}%
\bibitem [{\citenamefont {Castonguay}\ \emph {et~al.}(2022)\citenamefont
  {Castonguay}, \citenamefont {Kailasham}, \citenamefont {Wentworth},
  \citenamefont {Meredith}, \citenamefont {Khair},\ and\ \citenamefont
  {Zarzar}}]{Castonguay:22}%
  \BibitemOpen
  \bibfield  {author} {\bibinfo {author} {\bibfnamefont {A.~C.}\ \bibnamefont
  {Castonguay}}, \bibinfo {author} {\bibfnamefont {R.}~\bibnamefont
  {Kailasham}}, \bibinfo {author} {\bibfnamefont {C.~M.}\ \bibnamefont
  {Wentworth}}, \bibinfo {author} {\bibfnamefont {C.~H.}\ \bibnamefont
  {Meredith}}, \bibinfo {author} {\bibfnamefont {A.~S.}\ \bibnamefont {Khair}},
  \ and\ \bibinfo {author} {\bibfnamefont {L.~D.}\ \bibnamefont {Zarzar}},\
  }\bibfield  {title} {\enquote {\bibinfo {title} {Gravitational settling of
  active droplets},}\ }\href@noop {} {\  (\bibinfo {year} {2022})}\BibitemShut
  {NoStop}%
\bibitem [{\citenamefont {Stone}\ and\ \citenamefont
  {Samuel}(1996)}]{Stone:96}%
  \BibitemOpen
  \bibfield  {author} {\bibinfo {author} {\bibfnamefont {H.~A.}\ \bibnamefont
  {Stone}}\ and\ \bibinfo {author} {\bibfnamefont {A.~D.~T.}\ \bibnamefont
  {Samuel}},\ }\bibfield  {title} {\enquote {\bibinfo {title} {Propulsion of
  microorganisms by surface distortions},}\ }\href@noop {} {\bibfield
  {journal} {\bibinfo  {journal} {Phys. Rev. Lett.}\ }\textbf {\bibinfo
  {volume} {77}},\ \bibinfo {pages} {4102} (\bibinfo {year}
  {1996})}\BibitemShut {NoStop}%
\bibitem [{\citenamefont {Acrivos}\ and\ \citenamefont
  {Taylor}(1962)}]{Acrivos:62}%
  \BibitemOpen
  \bibfield  {author} {\bibinfo {author} {\bibfnamefont {A.}~\bibnamefont
  {Acrivos}}\ and\ \bibinfo {author} {\bibfnamefont {T.~D.}\ \bibnamefont
  {Taylor}},\ }\bibfield  {title} {\enquote {\bibinfo {title} {Heat and mass
  transfer from single spheres in {Stokes} flow},}\ }\href@noop {} {\bibfield
  {journal} {\bibinfo  {journal} {Phys. Fluids}\ }\textbf {\bibinfo {volume}
  {5}},\ \bibinfo {pages} {387--394} (\bibinfo {year} {1962})}\BibitemShut
  {NoStop}%
\bibitem [{\citenamefont {Feng}\ and\ \citenamefont
  {Michaelides}(1996)}]{Feng:96}%
  \BibitemOpen
  \bibfield  {author} {\bibinfo {author} {\bibfnamefont {Z.-G.}\ \bibnamefont
  {Feng}}\ and\ \bibinfo {author} {\bibfnamefont {E.~E.}\ \bibnamefont
  {Michaelides}},\ }\bibfield  {title} {\enquote {\bibinfo {title} {Unsteady
  heat transfer from a sphere at small peclet numbers},}\ }\href@noop {}
  {\bibfield  {journal} {\bibinfo  {journal} {J. Fluids Eng.}\ }\textbf
  {\bibinfo {volume} {118(1)}},\ \bibinfo {pages} {96--102} (\bibinfo {year}
  {1996})}\BibitemShut {NoStop}%
\bibitem [{\citenamefont {Pozrikidis}(1997)}]{Pozrikidis:97}%
  \BibitemOpen
  \bibfield  {author} {\bibinfo {author} {\bibfnamefont {C}~\bibnamefont
  {Pozrikidis}},\ }\bibfield  {title} {\enquote {\bibinfo {title} {Unsteady
  heat or mass transport from a suspended particle at low peclet numbers},}\
  }\href@noop {} {\bibfield  {journal} {\bibinfo  {journal} {J. Fluid Mech.}\
  }\textbf {\bibinfo {volume} {334}},\ \bibinfo {pages} {111--133} (\bibinfo
  {year} {1997})}\BibitemShut {NoStop}%
\bibitem [{\citenamefont {Abramowitz}\ and\ \citenamefont
  {Stegun}(1965)}]{Abramowitz:book}%
  \BibitemOpen
  \bibfield  {author} {\bibinfo {author} {\bibfnamefont {M.}~\bibnamefont
  {Abramowitz}}\ and\ \bibinfo {author} {\bibfnamefont {I.~A.}\ \bibnamefont
  {Stegun}},\ }\href@noop {} {\emph {\bibinfo {title} {Handbook of Mathematical
  Functions}}},\ \bibinfo {edition} {3rd}\ ed.\ (\bibinfo  {publisher}
  {Dover},\ \bibinfo {address} {New York, NY},\ \bibinfo {year}
  {1965})\BibitemShut {NoStop}%
\bibitem [{\citenamefont {Rednikov}\ \emph {et~al.}(1995)\citenamefont
  {Rednikov}, \citenamefont {Kurdyumov}, \citenamefont {Ryazantsev},\ and\
  \citenamefont {Velarde}}]{Rednikov:95}%
  \BibitemOpen
  \bibfield  {author} {\bibinfo {author} {\bibfnamefont {A.~Y.}\ \bibnamefont
  {Rednikov}}, \bibinfo {author} {\bibfnamefont {V.~N.}\ \bibnamefont
  {Kurdyumov}}, \bibinfo {author} {\bibfnamefont {Y.~S.}\ \bibnamefont
  {Ryazantsev}}, \ and\ \bibinfo {author} {\bibfnamefont {M.~G.}\ \bibnamefont
  {Velarde}},\ }\bibfield  {title} {\enquote {\bibinfo {title} {The role of
  time-varying gravity on the motion of a drop induced by marangoni
  instability},}\ }\href@noop {} {\bibfield  {journal} {\bibinfo  {journal}
  {Phys. Fluids}\ }\textbf {\bibinfo {volume} {7}},\ \bibinfo {pages}
  {2670--2678} (\bibinfo {year} {1995})}\BibitemShut {NoStop}%
\bibitem [{\citenamefont {Lippera}\ \emph
  {et~al.}(2020{\natexlab{a}})\citenamefont {Lippera}, \citenamefont {Morozov},
  \citenamefont {Benzaquen},\ and\ \citenamefont {Michelin}}]{Lippera:20}%
  \BibitemOpen
  \bibfield  {author} {\bibinfo {author} {\bibfnamefont {K.}~\bibnamefont
  {Lippera}}, \bibinfo {author} {\bibfnamefont {M.}~\bibnamefont {Morozov}},
  \bibinfo {author} {\bibfnamefont {M.}~\bibnamefont {Benzaquen}}, \ and\
  \bibinfo {author} {\bibfnamefont {S.}~\bibnamefont {Michelin}},\ }\bibfield
  {title} {\enquote {\bibinfo {title} {Collisions and rebounds of chemically
  active droplets},}\ }\href@noop {} {\bibfield  {journal} {\bibinfo  {journal}
  {J. Fluid Mech.}\ }\textbf {\bibinfo {volume} {886}},\ \bibinfo {pages}
  {1843--35} (\bibinfo {year} {2020}{\natexlab{a}})}\BibitemShut {NoStop}%
\bibitem [{\citenamefont {Farutin}\ \emph {et~al.}(2021)\citenamefont
  {Farutin}, \citenamefont {Rizvi}, \citenamefont {Hu}, \citenamefont {Lin},
  \citenamefont {Rafai},\ and\ \citenamefont {Misbah}}]{Farutin:21b}%
  \BibitemOpen
  \bibfield  {author} {\bibinfo {author} {\bibfnamefont {A.}~\bibnamefont
  {Farutin}}, \bibinfo {author} {\bibfnamefont {M.~S.}\ \bibnamefont {Rizvi}},
  \bibinfo {author} {\bibfnamefont {W.~F.}\ \bibnamefont {Hu}}, \bibinfo
  {author} {\bibfnamefont {T.~S.}\ \bibnamefont {Lin}}, \bibinfo {author}
  {\bibfnamefont {S.}~\bibnamefont {Rafai}}, \ and\ \bibinfo {author}
  {\bibfnamefont {C.}~\bibnamefont {Misbah}},\ }\bibfield  {title} {\enquote
  {\bibinfo {title} {A reduced model for a phoretic swimmer},}\ }\href@noop {}
  {\bibfield  {journal} {\bibinfo  {journal} {arXiv preprint arXiv:2112.12023}\
  } (\bibinfo {year} {2021})}\BibitemShut {NoStop}%
\bibitem [{\citenamefont {Lippera}\ \emph
  {et~al.}(2020{\natexlab{b}})\citenamefont {Lippera}, \citenamefont
  {Benzaquen},\ and\ \citenamefont {Michelin}}]{Lippera:20b}%
  \BibitemOpen
  \bibfield  {author} {\bibinfo {author} {\bibfnamefont {K.}~\bibnamefont
  {Lippera}}, \bibinfo {author} {\bibfnamefont {M.}~\bibnamefont {Benzaquen}},
  \ and\ \bibinfo {author} {\bibfnamefont {S.}~\bibnamefont {Michelin}},\
  }\bibfield  {title} {\enquote {\bibinfo {title} {Alignment and scattering of
  colliding active droplets},}\ }\href@noop {} {\bibfield  {journal} {\bibinfo
  {journal} {Soft Matter}\ } (\bibinfo {year}
  {2020}{\natexlab{b}})}\BibitemShut {NoStop}%
\bibitem [{\citenamefont {Chen}\ \emph {et~al.}(2021)\citenamefont {Chen},
  \citenamefont {Chong}, \citenamefont {Liu}, \citenamefont {Verzicco},\ and\
  \citenamefont {Lohse}}]{Chen:21}%
  \BibitemOpen
  \bibfield  {author} {\bibinfo {author} {\bibfnamefont {Y.}~\bibnamefont
  {Chen}}, \bibinfo {author} {\bibfnamefont {K.~L.}\ \bibnamefont {Chong}},
  \bibinfo {author} {\bibfnamefont {L.}~\bibnamefont {Liu}}, \bibinfo {author}
  {\bibfnamefont {R.}~\bibnamefont {Verzicco}}, \ and\ \bibinfo {author}
  {\bibfnamefont {D.}~\bibnamefont {Lohse}},\ }\bibfield  {title} {\enquote
  {\bibinfo {title} {Instabilities driven by diffusiophoretic flow on catalytic
  surfaces},}\ }\href@noop {} {\bibfield  {journal} {\bibinfo  {journal} {J.
  Fluid Mech.}\ }\textbf {\bibinfo {volume} {919}},\ \bibinfo {pages} {A10}
  (\bibinfo {year} {2021})}\BibitemShut {NoStop}%
\bibitem [{\citenamefont {Desai}\ and\ \citenamefont
  {Michelin}(2021)}]{Desai:21}%
  \BibitemOpen
  \bibfield  {author} {\bibinfo {author} {\bibfnamefont {N.}~\bibnamefont
  {Desai}}\ and\ \bibinfo {author} {\bibfnamefont {S.}~\bibnamefont
  {Michelin}},\ }\bibfield  {title} {\enquote {\bibinfo {title} {Instability
  and self-propulsion of active droplets along a wall},}\ }\href@noop {}
  {\bibfield  {journal} {\bibinfo  {journal} {Phys. Rev. Fluids}\ }\textbf
  {\bibinfo {volume} {6}},\ \bibinfo {pages} {114103} (\bibinfo {year}
  {2021})}\BibitemShut {NoStop}%
\bibitem [{\citenamefont {Picella}\ and\ \citenamefont
  {Michelin}(2022)}]{Picella:22}%
  \BibitemOpen
  \bibfield  {author} {\bibinfo {author} {\bibfnamefont {F.}~\bibnamefont
  {Picella}}\ and\ \bibinfo {author} {\bibfnamefont {S.}~\bibnamefont
  {Michelin}},\ }\bibfield  {title} {\enquote {\bibinfo {title} {Confined
  self-propulsion of an isotropic active colloid},}\ }\href@noop {} {\bibfield
  {journal} {\bibinfo  {journal} {J. Fluid Mech.}\ }\textbf {\bibinfo {volume}
  {933}} (\bibinfo {year} {2022})}\BibitemShut {NoStop}%
\bibitem [{\citenamefont {Desai}\ and\ \citenamefont
  {Michelin}(2022)}]{Desai:22}%
  \BibitemOpen
  \bibfield  {author} {\bibinfo {author} {\bibfnamefont {N.}~\bibnamefont
  {Desai}}\ and\ \bibinfo {author} {\bibfnamefont {S.}~\bibnamefont
  {Michelin}},\ }\bibfield  {title} {\enquote {\bibinfo {title} {Steady state
  propulsion of isotropic active colloids along a wall},}\ }\href@noop {}
  {\bibfield  {journal} {\bibinfo  {journal} {Phys. Rev. Fluids}\ }\textbf
  {\bibinfo {volume} {7}},\ \bibinfo {pages} {100501} (\bibinfo {year}
  {2022})}\BibitemShut {NoStop}%
\bibitem [{\citenamefont {Hokmabad}\ \emph {et~al.}(2022)\citenamefont
  {Hokmabad}, \citenamefont {Agudo-Canalejo}, \citenamefont {Saha},
  \citenamefont {Golestanian},\ and\ \citenamefont {Maass}}]{Hokmabad:22}%
  \BibitemOpen
  \bibfield  {author} {\bibinfo {author} {\bibfnamefont {B.~V.}\ \bibnamefont
  {Hokmabad}}, \bibinfo {author} {\bibfnamefont {J.}~\bibnamefont
  {Agudo-Canalejo}}, \bibinfo {author} {\bibfnamefont {S.}~\bibnamefont
  {Saha}}, \bibinfo {author} {\bibfnamefont {R.}~\bibnamefont {Golestanian}}, \
  and\ \bibinfo {author} {\bibfnamefont {C.~C.}\ \bibnamefont {Maass}},\
  }\bibfield  {title} {\enquote {\bibinfo {title} {Chemotactic self-caging in
  active emulsions},}\ }\href@noop {} {\bibfield  {journal} {\bibinfo
  {journal} {PNAS}\ }\textbf {\bibinfo {volume} {119}},\ \bibinfo {pages}
  {e2122269119} (\bibinfo {year} {2022})}\BibitemShut {NoStop}%
\end{thebibliography}%
\end{document}